\documentclass[twocolumn, tighten]{aastex63}

\definecolor{my_color}{HTML}{3a18b1}

\definecolor{new_color}{HTML}{CF0000}
\definecolor{new_green}{HTML}{008000}
\definecolor{new_black}{HTML}{000000}
\usepackage{xfrac}
\usepackage{mathtools}
\usepackage{amsmath}
\usepackage{verbatim}
\usepackage{longtable}

\newcommand\bmaroon{\textcolor{new_black}}

\newcommand\Kepler{\textit{Kepler}}
\newcommand\microhertz{\ensuremath{\mu \rm{Hz}}}

\newcommand\Spitzer{\textit{Spitzer}}
\newcommand\TESS{\textit{TESS}}
\newcommand\thisstar{HAT-P-7}
\newcommand\thisplanet{HAT-P-7~b}

\received{ }
\revised{ }
\accepted{ }
\shorttitle{Is HAT-P-7 \textup{b}'s Atmosphere Varying?}
\shortauthors{Lally \& Vanderburg}

\begin{document}

\title{Reassessing the Evidence for Time Variability in the Atmosphere of the Exoplanet HAT-P-7 b}

\correspondingauthor{Maura Lally}
\email{ml2289@cornell.edu}

\author[0000-0002-4443-6725]{Maura Lally}
\affiliation{Department of Astronomy, Cornell University, Ithaca, NY 14853, USA}
\affiliation{Northwestern University, Evanston, IL 60208, USA}
\affiliation{University of Texas at Austin, Austin, TX 78712, USA}

\author[0000-0001-7246-5438]{Andrew Vanderburg}
\affiliation{Department of Physics and Kavli Institute for Astrophysics and Space Research, Massachusetts Institute of Technology, Cambridge, MA 02139, USA}
\affiliation{University of Wisconsin-Madison, Madison, WI 53706, USA}
\affiliation{University of Texas at Austin, Austin, TX 78712, USA}
\affiliation{NASA Sagan Fellow}



\begin{abstract}

We reassess the claimed detection of variability in the atmosphere of the hot Jupiter \thisplanet, reported by \citet{armstrong2016}.  Although astronomers expect hot Jupiters to have changing atmospheres, variability is challenging to detect. 
We looked for time variation in the phase curves of \thisplanet\ in \Kepler\ data using similar methods to \citet{armstrong2016}, and identified apparently significant variations similar to what they found. Numerous tests show the variations to be mostly robust to different analysis strategies. 
\bmaroon{However, when we injected unchanging phase curve signals into the light curves of other stars and searched for variability, we often saw similar levels of variations as in the \thisstar\ light curve. Fourier analysis of the \thisstar\ light curve revealed background red noise from stellar supergranulation on timescales similar to the planet's orbital period. Tests of simulated light curves with the same level of noise as \thisstar's supergranulation show that this effect alone can cause the amplitude and phase offset variability we detect for \thisplanet. Therefore, the apparent variations in \thisplanet's atmosphere could instead be caused by non-planetary sources, most likely photometric variability due to supergranulation on the host star. }
\end{abstract}


\keywords{Exoplanets, Exoplanet atmospheric variability }


\section{Introduction}
\label{sec:intro}

Due to significant exoplanet research efforts in recent decades, there exists a wealth of observational data of transiting exoplanets. Optical telescopes such as the \Kepler\ Space Telescope and the Transiting Exoplanet Survey Satellite (\TESS) and infrared missions like the \Spitzer\ Space Telescope have regularly observed light curves of stars with transiting exoplanets, collecting years of photometric data (e.g. \citealt{esteves2}, \citealt{wong_TESS}, \citealt{wong_spitzer}). In addition, the \textit{Hubble Space Telescope} has collected spectroscopic data in order to learn more about the atmospheric composition of exoplanets \bmaroon{(e.g. \citealt{tinetti2007}, \citealt{Kreidberg2014}, \citealt{knutson2014}, \citealt{sing2016}, \citealt{barstow2017}, \citealt{skaf}, \citealt{pancet2019}, \citealt{pancet2021}, \citealt{foote2022}; for an overview, see \citealt{kreidberg_review2018}).}

Phase curves can be a valuable tool for analyzing exoplanet atmospheres.
A phase curve is a light curve showing the flux variations of a host star over the course of a planet's orbit, including the planet's transit and secondary eclipse. The shape of the phase curve depends on various properties of the planet, including atmospheric characteristics \citep{esteves}. In particular, thermal emission (heating of the planet's atmosphere) and reflected light (light from the host star reflected off the planet's atmosphere) are atmospheric processes which contribute to variations in the shape and amplitude of the phase curve. The effects of reflected light are most evident in optical observations such as those made by \Kepler, while thermal emission dominates longer wavelengths, including the IR observations from \Spitzer\ (see \citealt{shporer} for a comprehensive review of the topic). 


\bmaroon{Although time variability in exoplanet atmospheres has been researched and theorized about for some time now \citep{Rauscher, agol2009, Kawahara, Vidotto}, it remains a relatively new frontier.} Just like we experience changing weather patterns here on Earth, the atmospheres of exoplanets should also undergo changes on relatively short timescales. Detecting atmospheric variability of exoplanets through analysis of phase curves is an exciting prospect, but it comes with significant challenges. It is challenging to even detect an exoplanet atmosphere in the first place, so trying to measure small changes in such small signals increases the difficulty even further. In general, successfully detecting atmospheric variability requires very high signal-to-noise observations, taken consistently over a long period of time \citep{hidalgo}. 

Despite the challenges associated with detecting variability in the atmospheres of exoplanets, there have been several possible detections. Variability has been claimed in two exoplanets using reflected-light observations from \Kepler\ data: by \citet{armstrong2016} studying the planet \thisplanet, and \citet{jackson}, studying the planet \Kepler-76 b. Both of these studies found strong variations in the longitude of the brightest/most reflective point in the planets' atmospheres. \citet{armstrong2016} found that the brightest longitude of \thisplanet\ varies by as much as $\pm 41^{\circ}$, while \citet{jackson} found variations of up to $ \pm 49^{\circ}$ \bmaroon{for \Kepler-76 b}. Meanwhile, \citet{bell} reported changes in the longitude of the hottest point in the atmosphere of WASP 12 b in infrared observations from \Spitzer.   Most detections of atmospheric variability come from studies focusing on hot Jupiters (\citealt{armstrong2016}, \citealt{jackson}, \citealt{bell}, \citealt{Wilson2021})--a class of massive exoplanets with short orbital periods, making them ideal for this type of analysis--but it is possible to search for variations in the atmospheres of smaller planets as well. In particular, \citet{demory55cnc} have detected variability in the phase curve of the hot super-Earth 55 Cnc e, \bmaroon{and \citet{tamburo} later identified variability in the secondary eclipse depth of that planet.} 




Since the first claimed detections of atmospheric variability, there has been theoretical work to try to understand and explain these results. \citet{komacek} concluded that hot Jupiters may have time variability large enough to observe: their simulations predict that the hottest part of the atmspheres of hot Jupiters should show variability of up to $ \pm3^{\circ}$ from the time-averaged phase offset. This $ \pm3^{\circ}$ variability is significantly smaller than reported variability \citep{armstrong2016, jackson, bell}, implying that some other processes may be at play.  \citet{rogers} showed that strong magnetic fields can drive even larger shifts in the position of the hottest part of the planet's atmosphere, up to tens of degrees in orbital phase. Another possibility is that the large variations in reflected light phase curves from \Kepler\ are due to the movements of reflective clouds \citep{parmentier2016, parmentier2020, roman2020}, rather than movements of the hottest part of the atmosphere on which \citet{rogers} and \citet{komacek} focused. More studies of atmospheric variability are needed to understand which, if any, of these processes may be taking place in exoplanet atmospheres.

In this paper, we aim to reassess the evidence for the \citet{armstrong2016} detection of atmospheric variability due to cloud movements on \thisplanet. In particular, \citet{armstrong2016} claimed to detect statistically significant changes in the longitude of the peak of the phase curve of \thisplanet\ over time. Using the same \Kepler\ data and similar modeling methods, we detect similar variations in phase offset. These variations initially appear to be statistically significant and robust to different analysis strategies. However, injection/recovery tests yield variations of similar strength in known non-variable sources, calling into question the interpretation of the \citet{armstrong2016} result\bmaroon{, and we identify photometric variability due to supergranulation on the host star as a plausible explanation for the apparent variations in \thisplanet's phase curve.} Our paper is organized as follows: In Section \ref{sec:kepler data collection} we discuss the collection of the \Kepler\ light curve data. In section \ref{analysis}, we describe our analysis techniques in detail, \bmaroon{as well as a series of tests to determine the robustness of our results to different analysis choices. In Section \ref{results}, we describe the results of our analysis and compare them to the results of \citet{armstrong2016}.} \bmaroon{In Section \ref{causes}, we discuss several tests exploring possible causes of spurious variability signals.} In Section \ref{sec:discussion}, we discuss the implications of our tests, and we conclude in Section \ref{sec:conclusions}.
 
 \begin{figure*}[ht!]
 \centering
 	\includegraphics[width=6.5in]{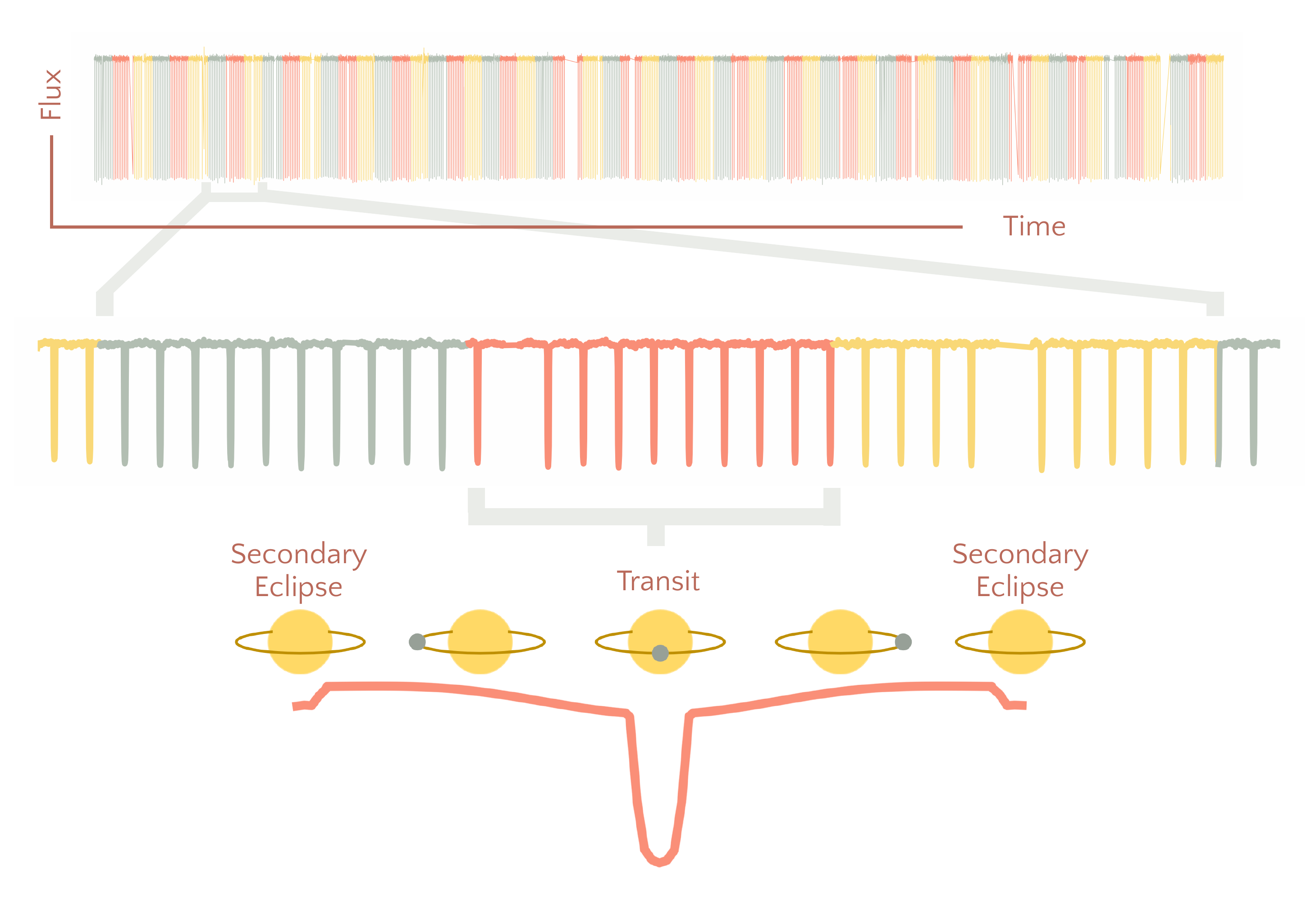}
     \caption{Overview of our analysis. \textit{Top:} The full \Kepler\ light curve of \thisstar, split into 60 consecutive bins of 10 orbits of \thisplanet, each shown as a different color/stripe. The transits of \thisplanet\ are so frequent that they blend together in this panel. \textit{Middle:} Close-up view of about 70 days of \Kepler\ data. In this panel, the transits of \thisplanet\ appear as a regular comb of dips (except for gaps in \Kepler's observations). \textit{Bottom:} An example planetary phase curve, shown alongside a schematic showing the relative position of the star and planet at several points in the orbit. We analyze each segment of 10 planet orbits (shown as different colors in the top two panels) to model a phase curve similar to that shown here.}
     \label{fig:schematic}
 \end{figure*}

\section{Kepler Data Collection}
\label{sec:kepler data collection}
\Kepler\ is an 0.95 meter space telescope launched into an Earth-trailing orbit with the primary goal of finding exoplanets using the transit method. During its operational period, \Kepler's mirror focused starlight onto its photometer, an array of CCDs that repeatedly imaged a roughly 110 square degree field of view and measured brightness variations in hundreds of thousands of stars. Before the failure of two reaction wheels ended its primary mission, \Kepler\ observed a field near the constellation Cygnus for over four years with a resolution of 4 arcsec/pixel \citep{borucki2010}. The spacecraft rotated by 90 degrees every 90 days in order to keep the solar panels aimed at the sun; as a result, \Kepler\ light curves are divided into 90 day quarters. The rotation of the spacecraft resulted in slow instrumental drifts in the data related to the cooling and heating of different parts of the telescope. Due to bandwidth limitations, data were partially analyzed on board, and only the highest priority data were transmitted back to Earth. Because the antenna was not remotely steerable, the spacecraft had to be reoriented in order to transmit data back to Earth, which involved interrupting a day of data collection for each communication. The \Kepler\ Science Operations Center at NASA Ames Research Center received raw \Kepler\ data, calibrated the images, produced light curves, removed systematic errors, and performed transit searches and candidate validation to verify exoplanet detections. We downloaded the \Kepler\ light curve of \thisplanet\ from the Mikulksi Archive for Space Telescopes (MAST) using the publicly available Lightkurve software package \citep{lightkurve}. 




\section{Kepler Data Analysis} \label{analysis}

\subsection{Baseline Analysis}\label{baseline}
After downloading the \Kepler\ light curve of \thisplanet, we performed analysis to detect and quantify variability in its phase curve. Our analysis consisted of two main steps: light curve flattening/removal of long-timescale variability, and MCMC modeling to determine the most probable phase curve parameters. 

\subsubsection{Removing Long-term Variability}

\Kepler\ light curves often show slow brightness variations on timescales of days to months. These variations can either be astrophysical (such as those caused by starspots coming in and out of view as the star rotates, e.g. \citealt{basri2013}) or instrumental (due to effects like the shifting position of the star on the detector due to differential velocity abberation, e.g. \citealt{jvc2016}). The long-term variations can be significantly larger than the amplitude of the phase curve signals we wish to study, so we must remove them before proceeding.

We started with the \Kepler\ light curves of \thisstar\ processed by the \Kepler\ team's Pre-search Data Conditioning (PDC) systematics correction software \citep{stumpe1, smith, stumpe2}. \Kepler\ observed \thisstar\ in both its standard ``long-cadence'' mode, where coadded images with exposure times of 29.4 minutes were downlinked from the spacecraft, and also in ``short-cadence'' mode, were coadded images with exposure times of 58 seconds were downloaded. Since we are primarily interested in light curve features on the timescale of \thisplanet's orbital period ($2.2$ days), the time resolution of the long-cadence light curve was sufficient for our analysis, and we chose to use it instead of the short-cadence light curve to speed up our computations. 

We fit the light curve with a basis spline, which is a piecewise series of cubic polynomials fit to sections of the light curve. The individual cubic polynomial fits are constrained so that the final spline curve is continuous and differentiable. We used a ``knot spacing'' of $2.205$ days (matching the orbital period of the planet) to prevent any attenuation of signals at the planet's orbital period by the spline. When performing the spline fit, we masked points within 2.5 hours of the times of transits and secondary eclipses. We prevented outlier data points from improperly affecting the spline fit by iteratively fitting the spline to the light curve, identifying the largest outliers, masking these points, and re-fitting the spline until convergence.  This process is illustrated in Figure 3 of \citet{vj14}. We divided the light curve by the best-fit spline to remove the long-timescale variability. This approach is similar to that taken by \citet{armstrong2016}, who fit a cubic polynomial in windows across the light curve.

\subsubsection{Fitting the Phase Curve with MCMC}\label{sec:mcmcfitting}
\begin{figure*}[ht!]
\centering
	\includegraphics[width=6.5in]{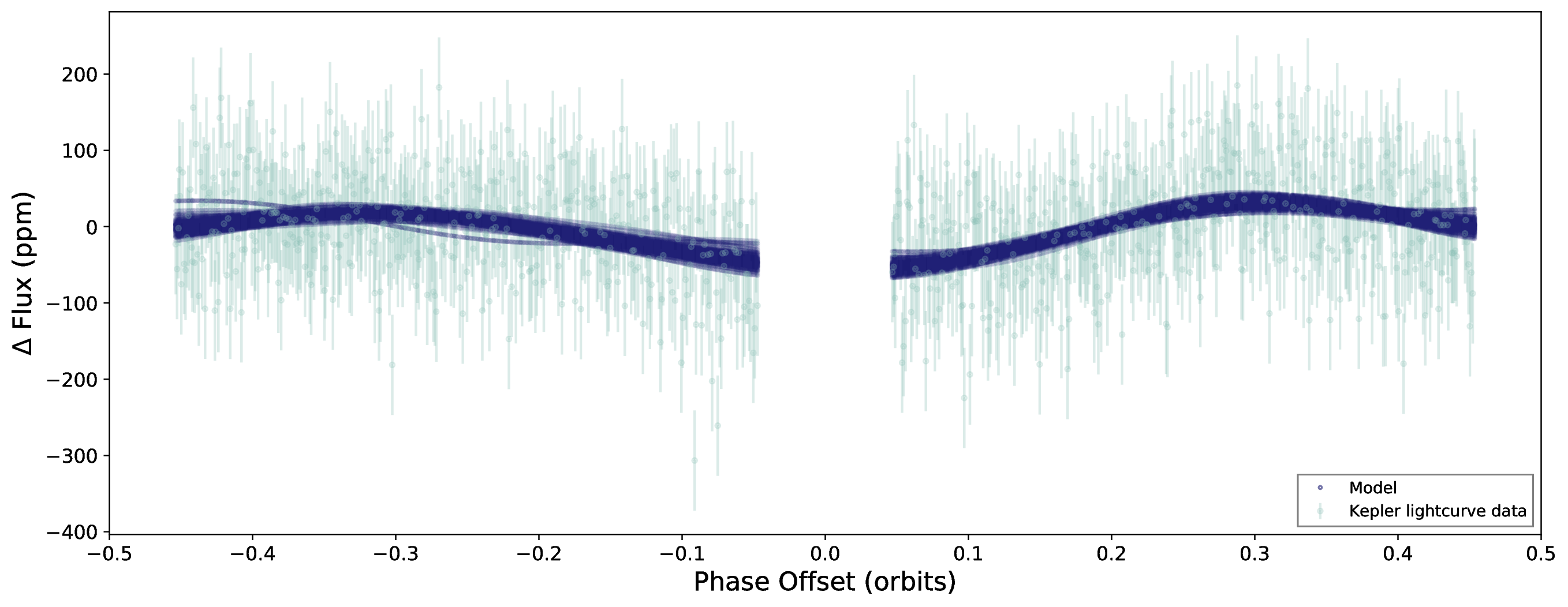}
    \caption{An example phase curve fit. The green points show about ten phase-folded orbits of \Kepler\ data (excluding the transit and secondary eclipses), while the blue lines show the models corresponding to 100 draws from our MCMC posterior probability distributions of our sinusoidal model fit to the data. We perform a similar fit every ten orbits observed by \Kepler\ to detect any variations in the phase curve over time.}
    \label{fig:fitplot}
\end{figure*}
After flattening the light curve, we fitted a model to the light curve to measure the best-fit values and uncertainties of parameters describing \thisplanet's phase curve. We split the full four-year light curve into approximately 60 different segments, each about 22 days in length (or 10 orbits of the planet \thisplanet). We chose this segment length following \citet{armstrong2016}; fitting 10 planet orbits is a good balance between high enough time resolution to probe atmospheric variability, and fitting enough data to confidently detect the phase curve in each segment and minimize the influence of any remaining instrumental systematics which might still be present. 
We removed points within 30 minutes before and after the duration of transits and secondary eclipses from the light curve to avoid calculating computationally expensive transit and eclipse models at each step. \bmaroon{We modeled the phase curve in each time segment using Easy Differential Evolution Markov Chain Monte Carlo (EDMCMC\footnote{For more detail, see https://github.com/avanderburg/edmcmc}), a sampling routine which makes use of the differential evolution MCMC algorithm of \citet{terBraak}, which in our case achieves better convergence than conventional MCMC. We utilized the jump scheme described by \citet{terBraak} where every 10 steps, the MCMC attempts a larger jump than usual. This provides better convergence when there are bimodalities in the posterior distributions.} We fit each light curve segment with a simple sinusoidal flux model $F$ given by: 
\begin{multline}\label{model}
        F = c+a\sin\left(\frac{2\pi t}{P}+2\phi \pi\right) +k\sin\left(\frac{2\pi t}{P/2}+2\phi_2 \pi\right) \\
        +h\sin\left(\frac{2\pi t}{P/3}+2\phi_3 \pi\right)
\end{multline}
\noindent \bmaroon{where $a$, $k$, and $h$ are the amplitudes of sine functions at the orbital period and its second and third harmonics, $\phi$, $\phi_2$, and $\phi_3$ are phase offsets of the three sinusoids, and $t$ is the time of each \Kepler\ observation. The model fits for eight free parameters: period $P$ (the orbital period of \thisplanet, constrained with an informative Gaussian prior to be $2.204735417 \pm  4.3\times10^{-8}$ days, \citealt{thompson2018}), constant flux offset $c$, and combinations of amplitudes and phases $x$, $y$, $x_2$, $y_2$, $x_3$, $y_3$. 
We define the combined amplitude and phase parameters as:}

\begin{eqnarray}
x \equiv \sqrt{a}\cos{\phi}\\
y \equiv \sqrt{a}\sin{\phi}\\
x_2 \equiv \sqrt{k}\cos{\phi_2}\\
y_2 \equiv \sqrt{k}\sin{\phi_2}\\
x_3 \equiv \sqrt{h}\cos{\phi_3}\\
y_3 \equiv \sqrt{h}\sin{\phi_3}
\end{eqnarray}

\noindent This model is equivalent to that of \citet{armstrong2016}, except that they also model the secondary eclipses, while we exclude these regions from our fit.


We used a $\chi^2$ log-likelihood function, allowing the typical uncertainty of the \Kepler\ flux measurements to vary as a free parameter. In particular, our log-likelihood function, $\log{\mathcal{L}}$, is given by: 
\begin{equation}
    \log{\mathcal{L}} = - \sum_i \left[ \frac{(y_i - F_i)^2}{2\sigma_K^2} + \log{\sigma_K}\right]
\end{equation}
\noindent where $y_i$ are the individual \Kepler\ flux measurements, $F_i$ are the individual evaluations of our model (Equation \ref{model}) and $\sigma_K$ is a free parameter representing the uncertainty of each \Kepler\ flux measurement. In total, we explored eight free parameters with our MCMC with 100 walkers evolved for 5000 steps, discarding the first 2000 as burn-in. We performed a similar fit for each of the 10-orbit light curve segments. The results for a typical phase curve fit are shown in Figure \ref{fig:fitplot}. 

After fitting each light curve segment with MCMC, we extracted the most likely parameters and uncertainties for the phase ($\phi$) and amplitude ($a$) of the sine at the planet's orbital period and collected them into time series. These are the model parameters that \citet{armstrong2016} found to show time variability, although after some testing they concluded that the variations in amplitude ($a$) were likely spurious and only the variations in phase ($\phi$) were likely due to atmospheric variability. 

\bmaroon{We tested the convergence of our sampler chains using the Gelman-Rubin diagnostic described in \citet{gelman_rubin}. We found that all of our chains for each time segment had a Gelman-Rubin statistic less than 1.2, all but one were less than 1.1, and all but two (99.6\%) of the chains had a Gelman-Rubin statistic less than 1.05. While a handful of these chains did not achieve convergence according to some of the stricter definitions \citep[e.g.][]{vats2021revisiting}, the unconverged chains were the higher-order harmonic signals that we treat as nuisance parameters (see Section \ref{sec:model harmonics}). We inspected these cases and found that the convergence of the parameters describing the phase and amplitude at the planet's orbital period was not affected. } 

\subsection{Testing the Robustness of our Analysis}\label{robustness}

    

In the previous subsection, we described the different steps we took to analyze the \Kepler\ data and measure phase curve parameters. Throughout our analysis, we made choices that could in principle affect the results of our analysis.  Here, we describe tests we performed to determine how robust our analysis is to these different choices. In particular, we compare measurements of the amplitude ($a$)  and phase ($\phi$)  of the sine at the planet's orbital period for each different analysis. Since these are the parameters found by \citet{armstrong2016} to vary in their analysis, measuring similar values for these parameters despite different analysis choices indicates that our analysis is robust to these different analysis strategies.

\subsubsection{Data Processing/Systematics Correction}
\label{sec:processing data}
\begin{figure*}[ht!]
    \centering
    	\includegraphics[width=5.5in]{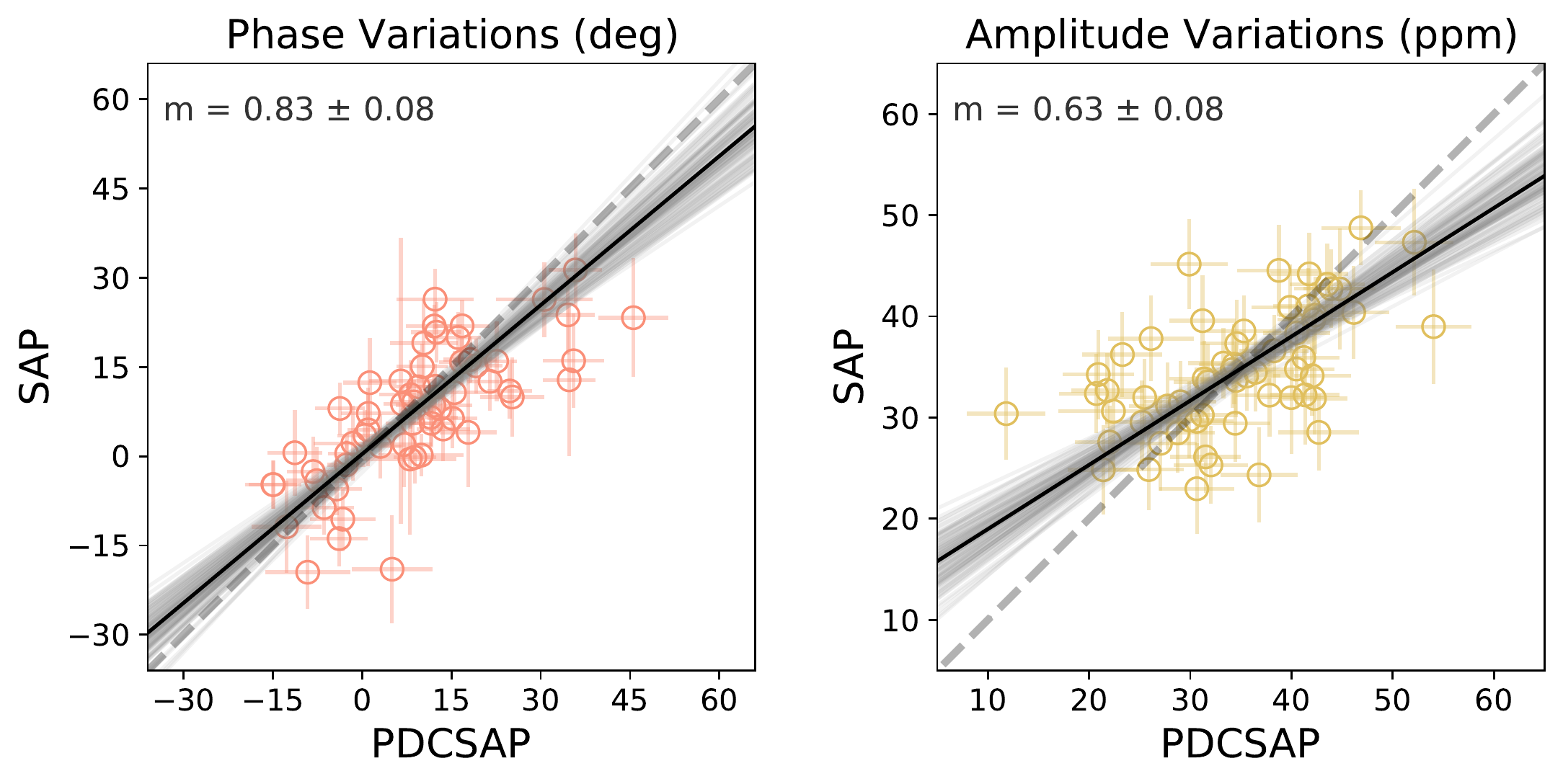}
        \caption{Comparison between phase offset variations (\textit{Left}) and amplitude variations (\textit{Right}) resulting from the use of \Kepler\ data processed via Simple Aperature Photometry (SAP) vs. Pre-search Conditioning Simple Aperature Photometry (PDCSAP). The \bmaroon{dashed gray diagonal lines in both panels show a one-to-one slope, while the solid black line shows the slope from a fit accommodating the error bars in both axes. The fitted slope and uncertainty appears on the top left of each plot. Our analysis shows some minor systematic differences between phases and amplitudes we measure from SAP and PDCSAP data, but qualitatively, the results are similar.} We chose to use PDCSAP data for our analysis, to ensure that our result did not depend on possible systematics present in the SAP data.}
        \label{fig:pdc_vs_sap}
\end{figure*}
The first analysis choice we made was whether to use the \Kepler\ data with a systematics correction or the raw un-processed light curve. The \Kepler\ archive stores both the raw light curves produced by Simple Aperture Photometry (SAP) as well as the post-processed Pre-search Data Conditioning Simple Aperture Photometry (PDCSAP) light curves. The PDCSAP light curves have been processed by the PDC module of the \Kepler\ pipeline \citep{stumpe1,smith,stumpe2}, which removes systematics and corrects for contamination from nearby sources. Usually, PDCSAP light curves are higher-quality than SAP light curves; however, the PDC processing is done in bulk to all targets observed by \Kepler\ and is not tuned to individual targets, leaving open the possibility that the algorithm could perform suboptimally on any given target. Also, \thisstar\ shows high-signal-to-noise photometric variability (i.e. the planetary transits), which in principle could hurt the quality of the systmatics correction\footnote{The \Kepler\ pipeline does automatically ignore transits of known planetary signals when performing the PDC correction, which should decrease the risk of failure on stars like \thisstar\ (J. Smith, priv. comm.).}. In fact, \citet{armstrong2016} chose to use the SAP light curve for their own analysis, likely to avoid such complications. 

We therefore tested to determine whether our choice to use the PDCSAP light curve instead of the unprocessed SAP light curve significantly affected the results of our MCMC fits. We ran our MCMC analysis on all 60 light curve segments in both the SAP light curve and PDCSAP light curve, with all other parameters and choices held identical, and compared each measurement of the phase ($\phi$) and amplitude ($a$) of the sine at the planet's orbital period. The phase and amplitude measurements for the PDCSAP and SAP light curves are compared in Figure \ref{fig:pdc_vs_sap}.  We found that the two different data processing methods described yielded \bmaroon{qualitatively} similar amplitude and phase measurements. \bmaroon{We quantified this correspondence by finding the best-fit linear relationship between the phases and amplitudes measured from the PDC and SAP light curves. We accounted for errors in both the SAP and PDC light curve measurements following the prescription in Equation 2 of \citet{press1992}. Although we found some evidence for slight systematic differences between the SAP and PDC datasets (with SAP yielding smaller phase offsets and amplitudes), largely the two datasets showed good correspondence.} For the the rest of our analysis, we use the PDCSAP processed data.

\subsubsection{Flattening/Removal of long-term variability}

    \begin{figure*}[ht!]
    \centering
    	\includegraphics[width=5.5in]{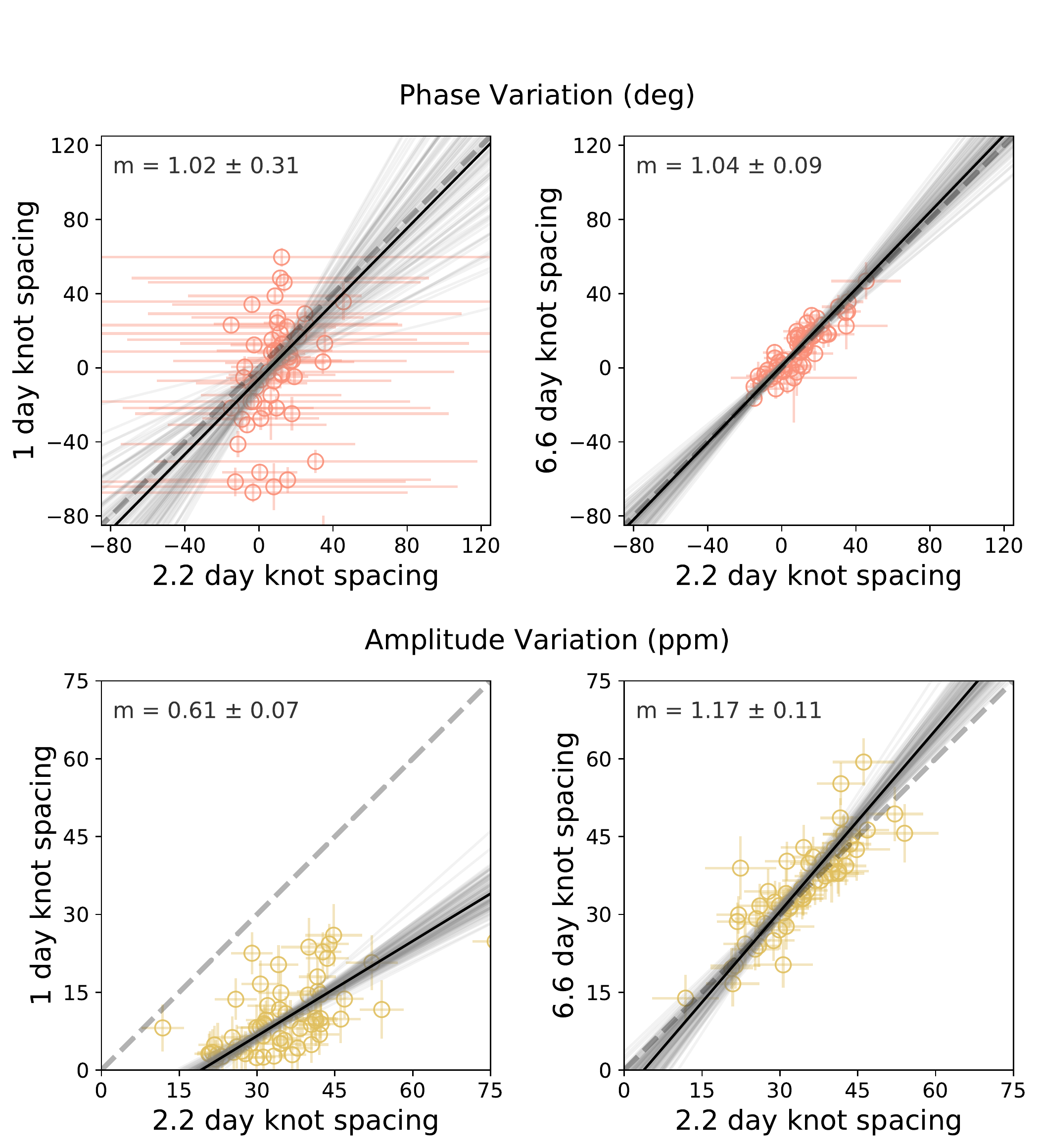}
        \caption{Comparison between phase (\textit{Top Row}) and amplitude variations (\textit{Bottom Row}) in \thisplanet's phase curve when we detrend the data using a basis spline with different knot spacings. \bmaroon{The dashed gray diagonal lines in each plot show a one-to-one slope, while the solid black line shows the slope from a fit accommodating the error bars in both axes. The fitted slope and uncertainty appears on the top left of each plot.} For aggressive splines with knot spacings shorter than duration of one \thisplanet\ orbit (\textit{Left Column}), the phase curve amplitude was significantly attenuated ,and the resulting small signal resulted in highly uncertain measurements of the phase of peak brightness. Knot spacing greater than or equal to one planet orbit (\textit{Right Column}) generally were consistent. We chose to use a spacing of one knot per orbit for our analysis to give our spline the most flexibility possible without attenuating the signal.}
        \label{fig:brk}
    \end{figure*}

We tested several different methods for flattening the light curve \bmaroon{to remove} long-term variability. First, we tested whether different knot spacings for our basis spline could affect our measurements of \thisplanet's phase curve parameters. The spacing of spline knots controls the aggression of the spline and its ability to model fast variations. The closer in time the spline knots are spaced, the more effectively the spline can model and remove unwanted stellar or instrumental variability, but also the greater risk that the spline will partially model and attenuate \thisplanet's phase curve signal.  We therefore tested a range of different spline knot spacings. Figure \ref{fig:brk} shows several comparisons of our measurements of the phase ($\phi$) and amplitude ($a$) of the sine at the planet's orbital period with different knot spacings.  We found that spline knot spacings shorter than the planet's orbital period significantly attenuated the phase curve amplitude (to the point of being undetectable, which resulted in large scatter in the measured phases), while spacings greater than or equal to the planet's orbital period all seemed to preserve the phase curve signal and gave similar results. We therefore opted to use a spacing equal to the planet's orbital period: the most flexible spline model possible that did not appear to attenuate the amplitude of the phase curve.  

\begin{figure*}[ht!]
    \centering
    	\includegraphics[width=5.5in]{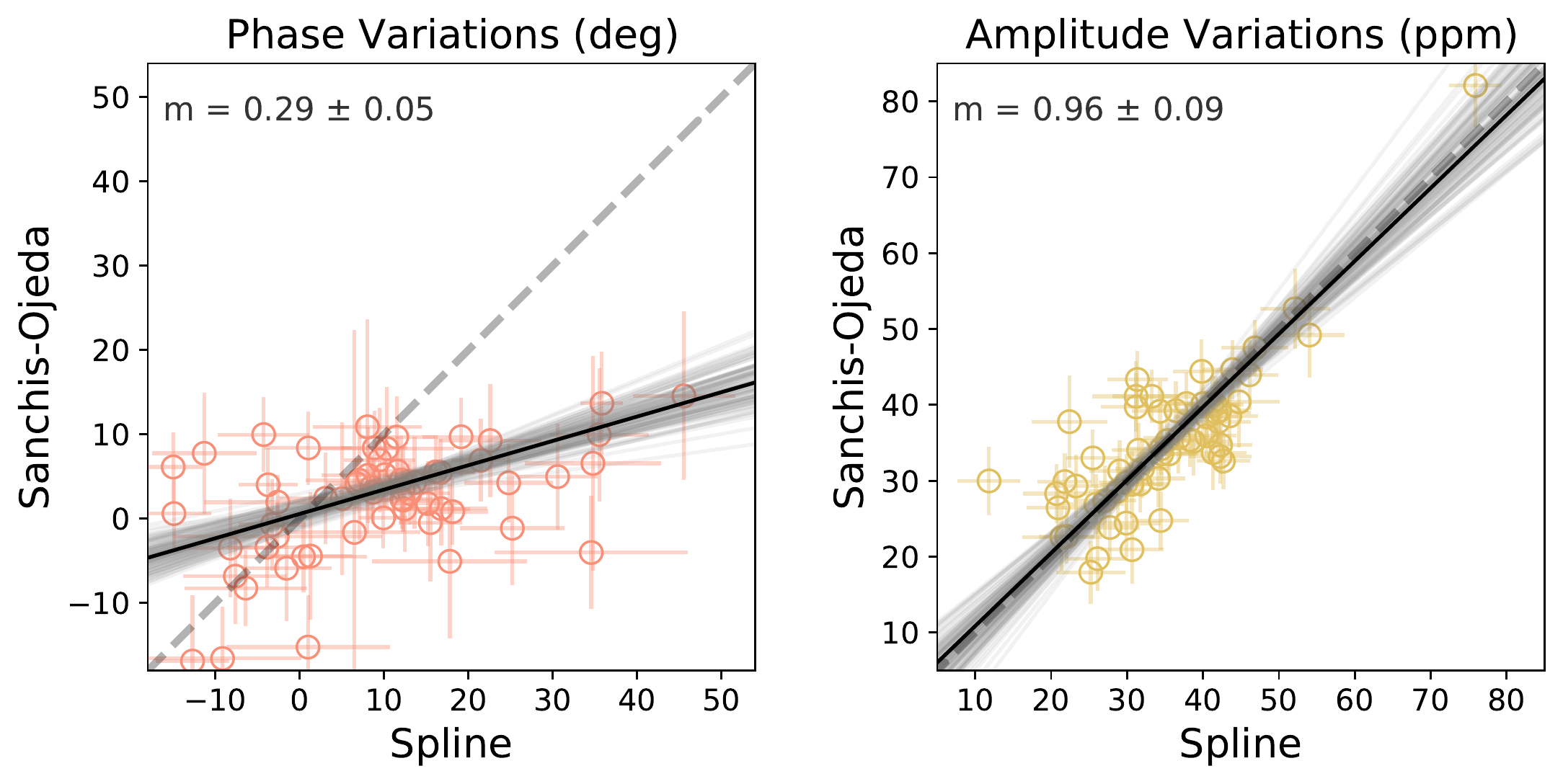}
        \caption{Phase (\textit{Left}) and amplitude (\textit{Right}) variations resulting from flattening the \Kepler\ light curve using the \citet{sanchis-ojeda} inspired filtering versus the basis spline method. \bmaroon{The dashed gray diagonal lines in each plot show a one-to-one slope, while the solid black line shows the slope from a fit accommodating the error bars in both axes. The fitted slope and uncertainty appears on the top left of each plot.} Both methods yield consistent measurements of the phase curve amplitude, but the \citet{sanchis-ojeda} inspired filtering results in smaller variations in the planet's orbital phase. It is unclear whether the smaller phase variations from the \citet{sanchis-ojeda} inspired method are due to better removal of systematic trends, or attenuation of real phase variations, so to avoid removing potentially real variability, we chose to use a basis spline to detrend the \Kepler\ data for our analysis.}
        \label{fig:so_v_spline}
\end{figure*}

We also tested another method for removing long-term trends from the light curve very similar to that proposed by \citet{sanchis-ojeda}. In this method, all in-transit data were masked (including points within a half hour before and after the transit duration), and a linear function of time was fit to all out-of-transit data, one orbit at a time. The best-fit linear function for each orbit was then divided from the light curve,
removing the long-term trends. When we tested this \citet{sanchis-ojeda} inspired method \bmaroon{and compared it to our spline results using the method described in Section \ref{sec:processing data}}, we found similar variations in the phase curve amplitude ($a$) compared to the values we measured when using spline detrending, but the phase variations ($\phi$) were substantially reduced compared to the spline detrending (see Figure \ref{fig:so_v_spline}). Evidently, the method used to model and remove low-frequency variations can significantly affect measurements of phase curve variability. It is not clear, however, whether the \citet{sanchis-ojeda} inspired detrending shows less phase variability because it suppresses real phase curve variations, or because it better models long-term trends and prevents the detection of spurious variability.  To make sure we do not inadvertently remove the variability signal we hoped to detect, we opt for the less aggressive basis-spline low-frequency removal approach.


\subsubsection{Model Harmonics}
\label{sec:model harmonics}
\begin{figure*}[ht!]
    \centering
    	\includegraphics[width=5.5in]{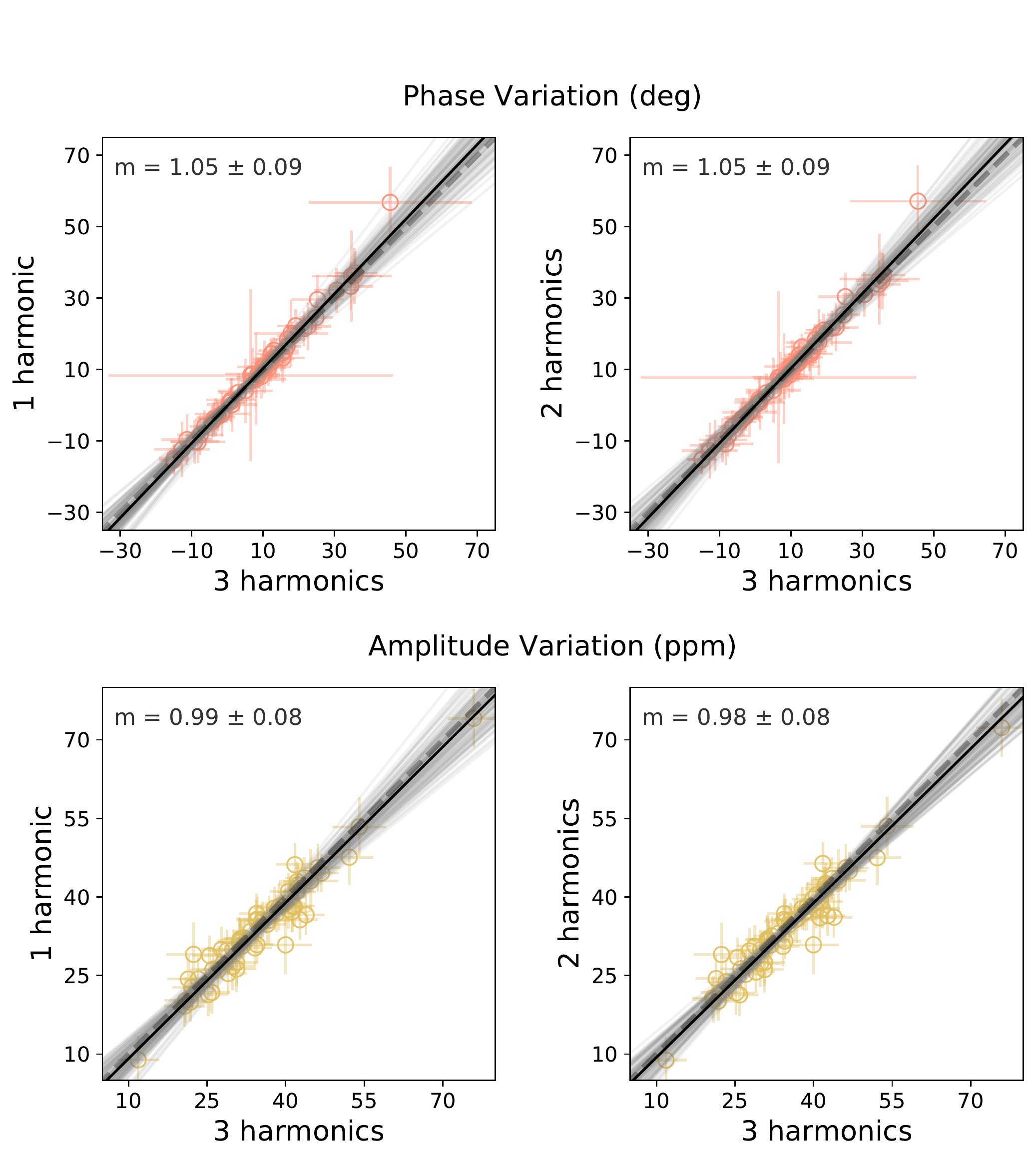}
        \caption{Comparison between phase (\textit{Top row}) and amplitude (\textit{Bottom Row}) variations for the sine at the planet's orbital period for our baseline model with three sinusoids compared with models including only one one (\textit{Left Column}) or two (\textit{Right Column}) sinusoids in the phase curve model. We note that all of the measurements plotted show the phase ($\phi$) and amplitude ($a$) of the sine at the planet's orbital period - the only differences between the measurements are the number of \textit{additional} sinusoid terms in the model. \bmaroon{The dashed gray diagonal lines in each plot show a one-to-one slope, while the solid black line shows the slope from a fit accommodating the error bars in both axes. The fitted slope and uncertainty appears on the top left of each plot.} The inclusion of more harmonics in our model did not cause significant change in our results for the phase and amplitude of the sine at the planet's orbital period. In order to maintain consistency with the methods of \citet{armstrong2016}, we used a model including three sinusoids for our analysis.}
        \label{fig:harm_comp}
    \end{figure*}

The last major analysis choice we made was which specific phase curve model to fit with MCMC. In particular, in our baseline analysis, we fit \thisplanet's phase curve with three sinusoids: one at the planet's orbital period and each of the second and third harmonics (half and one third of the orbital period). We tested to make sure that this particular choice of model does not significantly affect the measurements of the amplitude and phase of the sine at the planet's orbital period. We ran the MCMC fitting on sinusoidal models with two sinusoids (at the planet's orbital period and the second harmonic at half the orbital period) and only one sinusoid at the orbital period. Our measurements of the phase ($\phi$) and amplitude ($a$) of the sine at the planet's orbital period were not dependent on the number of harmonics included in the model. Figure \ref{fig:harm_comp} shows our measurements for these different MCMC models \bmaroon{using the linear fitting comparison method described in Section \ref{sec:processing data}}. In the end, we chose to use the model with three sinusoids to maintain consistency with the analysis of \citet{armstrong2016}.  

\subsection{Injection/Recovery Tests} 
\label{sec:injection/recovery}
In addition to performing tests to determine the robustness of our analysis to different choices or methods, we performed an end-to-end test of our analysis with injection/recovery tests. Our strategy was to inject a non-varying phase curve signal (based on the average phase curve of \thisplanet) into a number of other \Kepler\ light curves and to measure the phase curve parameters using the same analysis procedure as for \thisplanet. These tests allowed us to determine whether our analysis is susceptible to detecting spurious phase curve variations. If we detect statistically significant variations in the phase curves of these injected signals, they must be spurious because we know that the true (injected) signal's phase curve does not vary in time. \bmaroon{A similar test was also conducted by \citet{armstrong2016}, which led them to conclude that the variation they detected in phase curve amplitude could be spurious.}

We started by identifying a list of stars with similar properties to \thisstar. We searched for stars observed for the full \Kepler\ mission (17 quarters) with two criteria: either (a) stars within 0.05 magnitudes of \thisstar\ in \Kepler-band brightness and within 5\% (or 0.1 $R_\odot$) in size, or (b) stars within 0.2 magnitudes of \thisstar\ in \Kepler-band brightness, 10\% (or 0.2 $R_\odot$) in size, and 3\% (200 K) in effective temperature. A total of 28 stars satisfied at least one of these two criteria. We somewhat arbitrarily selected 10 of these stars to perform the injection/recovery tests. \bmaroon{We note that we did not solely choose stars with similar photometric variability characteristics to \thisstar; some of these other stars are quieter, and some are significantly noisier than \thisstar.}

We based the injected signal off of \thisplanet's average phase curve over the full four-year \Kepler\ mission.  We fitted analytic transit models \citep{MandelAgol:2002} to the transits and secondary eclipses, and modeled the average phase curve with a 100-segment piecewise linear function.  We then injected the transit+eclipse+phase curve model into each of the light curves of these other stars, ensuring that the injected phase curve signal was perfectly non-variable for the full four years of data.  We then processed, flattened, and modeled the injected curve for each other star in the same way as the original \thisplanet\ light curve. We collected phase and amplitude time series for the sine at the planet's orbital period for each injected star. \bmaroon{The results of this test are described in Section \ref{sec:assessing evidence for astrophysical variability}.}  

\section{Results} \label{results}
\begin{figure*}[ht!]
\centering
    \includegraphics[width=6.5in]{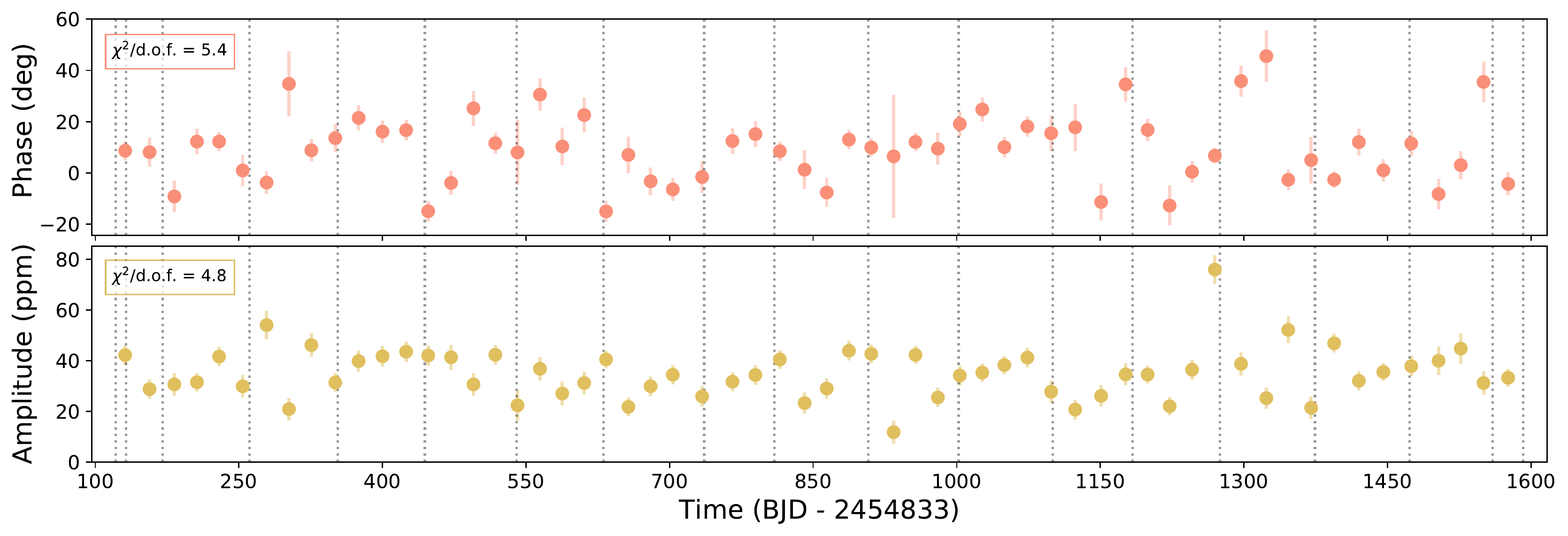}
    \caption{Measurements of the phase (\textit{Top}) and amplitude (\textit{Bottom}) of \thisplanet's phase curve over time. The vertical dotted lines show where the \Kepler\ light curve is broken into quarters. The reduced $\chi^2$, or $\chi^2$ per degree of freedom, is shown in the upper left of each plot. Although there are no clear patterns in the time evolution of the phase and amplitude of \thisplanet's phase curve, we find that there may be statistically significant variability based on a $\chi^2$ test.}
    \label{fig:ourresults}
\end{figure*}

\subsection{Measurements of variability in \thisplanet's Phase Offset and Amplitude}
\label{sec:measurements of hatp7b's phase variations and amplitude variations}

We show the measured phases ($\phi$) and amplitudes ($a$) of the sine at the orbital period of \thisplanet\ for our baseline analysis (described in Section \ref{baseline}) in Figure \ref{fig:ourresults} \bmaroon{and Table \ref{results_table}}. \bmaroon{Our results converge according to the Gelman-Rubin test described in Section \ref{baseline}}.  We appear to detect statistically significant changes in the phase and amplitude of \thisplanet's phase curve.  We find that the longitude of the peak of the phase curve varies by up to \bmaroon{$\pm$ 30} degrees, and the amplitude of the phase curve appears to vary by \bmaroon{$\pm 32$ ppm}. We quantify the statistical significance of any variations by calculating the $\chi^2$ statistic assuming a non-varying phase curve, which is defined as: 

\begin{equation}
    \chi^2 = \sum_i^N\left[\frac{{d_i - \bar d}}{\sigma_{d,i}}\right]^2
\end{equation}

\noindent where $N$ is the number of datapoints in the phase or amplitude time series, $d_i$ are the central values of the phase or amplitude measurement at each time, $\bar d$ are the predicted values of the phase/amplitude from our model (in this case, the mean value of these parameters since we assume the phase curve is not varying), and $\sigma_{d,i}$ are the uncertainties of each phase/amplitude measurement from our MCMC fit. We calculate $\chi^2$ values of \bmaroon{316} and \bmaroon{281} for the phase and amplitude time series, respectively. When we perform a $\chi^2$ test on these results, we find that it is exceedingly unlikely ($p \approx 0$ for both) that we would randomly measure such a large $\chi^2$ value, under the assumptions that the phase and amplitude are really constant, and that our uncertainties are properly estimated.

We also calculate $\chi^2/d.o.f.$, the $\chi^2$ value per degree of freedom, which is also called the reduced $\chi^2$ value. This metric is easier to interpret and compare between different datasets and time series, especially if they have different numbers of datapoints. We calculate $\chi^2/d.o.f.$ for our single-parameter constant model with:

\begin{equation}
    \chi^2/d.o.f = \frac{\chi^2}{N-M}
\end{equation}

\noindent where $N$ is the number of datapoints in the phase or amplitude time series, and $M$ is the number of parameters in our model (in this case, $M=1$, since we assume a constant model for the phase and amplitude time series). In cases where the model appropriately describes the data, and the uncertainties are estimated correctly, $\chi^2/d.o.f.$ should be close to 1. We measure a reduced $\chi^2$ value of 5.4 in the phase measurements and a reduced $\chi^2$ value of \bmaroon{4.8} for the amplitude measurements. \bmaroon{We confirmed that these high $\chi^2$ values are not due to an underestimate of the uncertainties on our individual \Kepler\ data points by calculating the $\chi^2$ for each of the 60 phase curve MCMC fits that we used to measure the phase offset and amplitude values. The $\chi^2 / d.o.f$ values for these phase curves have a median of $1.12$ and a standard deviation of $0.05$. Although this indicates slightly underestimated errors, it does not account for the large $\chi^2 / d.o.f$ that we find for the variations of phase offset and amplitude.}

Under ideal circumstances, the high values we measure for both $\chi^2$ and $\chi^2/d.o.f.$ could be evidence for variations in the atmosphere of \thisplanet. However, given the challenging nature of the measurement, some more scrutiny is necessary to conclude that the variations are indeed due to atmospheric variations. We describe additional tests to ascertain whether these apparent phase and amplitude variations are really likely to be astrophysical in Section \ref{sec:assessing evidence for astrophysical variability}.

\begin{deluxetable}{ccccc}[htbp]
\label{results_table}
\tabletypesize{\scriptsize}
\tablecolumns{1}
\tablecaption{Measurements of Phase and Amplitude for \thisplanet's Phase Curve \label{bigtable}}
\tablewidth{0pt}
\tablehead{
  \colhead{Time} & 
  \colhead{Phase ($\phi$)}     &
  \colhead{Uncertainty}     &
  \colhead{Amplitude($a$)}     &
  \colhead{Uncertainty}   \\[-0.4cm]
  \colhead{(BJD-2454833)} & 
  \colhead{degrees} &
  \colhead{degrees}  &
  \colhead{ppm}  & 
  \colhead{ppm} 
}
\startdata
131.3 & 9.0 & 3.8 & 42.1 & 3.8 \\
156.7 & 7.7 & 5.9 & 28.6 & 3.9 \\
182.6 & -9.0 & 6.2 & 30.7 & 4.5 \\
206.2 & 12.5 & 5.1 & 31.4 & 3.4 \\
229.4 & 12.6 & 3.8 & 41.0 & 3.8 \\
254.1 & 1.4 & 6.1 & 30.0 & 4.6 \\
279.0 & -3.7 & 4.4 & 54.1 & 5.5 \\
302.4 & 34.4 & 11.0 & 20.5 & 4.3 \\
325.8 & 8.1 & 4.3 & 47.2 & 4.8 \\
350.8 & 13.5 & 5.6 & 31.5 & 3.7 \\
375.2 & 21.4 & 4.9 & 39.9 & 4.0 \\
400.1 & 16.1 & 4.2 & 42.5 & 4.0 \\
424.8 & 16.8 & 4.0 & 43.8 & 3.8 \\
447.8 & -15.0 & 4.1 & 41.7 & 4.0 \\
471.7 & -3.4 & 4.5 & 43.4 & 4.8 \\
495.1 & 24.8 & 6.3 & 31.7 & 4.2 \\
517.9 & 11.7 & 4.1 & 42.0 & 4.0 \\
541.2 & 8.2 & 11.5 & 23.7 & 6.0 \\
564.6 & 31.7 & 6.4 & 35.8 & 4.8 \\
587.9 & 10.3 & 7.3 & 26.7 & 4.6 \\
610.8 & 22.6 & 6.6 & 31.5 & 4.5 \\
633.7 & -14.9 & 4.1 & 41.9 & 4.1 \\
657.0 & 6.8 & 7.1 & 21.5 & 3.6 \\
680.2 & -3.6 & 5.6 & 29.8 & 3.8 \\
703.4 & -6.3 & 4.5 & 34.8 & 3.7 \\
734.0 & -1.4 & 6.5 & 25.4 & 4.1 \\
765.7 & 12.7 & 5.1 & 31.6 & 3.6 \\
789.7 & 15.6 & 5.0 & 34.4 & 4.0 \\
815.2 & 8.7 & 3.9 & 39.8 & 3.6 \\
841.1 & 1.1 & 7.4 & 23.6 & 4.1 \\
864.2 & -7.5 & 5.7 & 28.7 & 4.1 \\
887.4 & 13.1 & 3.8 & 43.4 & 3.9 \\
910.7 & 9.6 & 3.7 & 42.2 & 3.8 \\
934.1 & 7.3 & 21.8 & 11.5 & 4.5 \\
956.9 & 12.2 & 3.6 & 42.3 & 3.6 \\
980.3 & 9.9 & 6.5 & 24.7 & 3.7 \\
1003.2 & 18.6 & 4.6 & 34.8 & 3.7 \\
1026.7 & 25.1 & 4.7 & 35.0 & 3.5 \\
1049.8 & 10.3 & 4.0 & 38.0 & 3.5 \\
1073.8 & 18.2 & 4.3 & 40.1 & 4.1 \\
1098.7 & 15.0 & 7.0 & 27.8 & 4.3 \\
1123.8 & 18.7 & 9.6 & 20.6 & 4.2 \\
1150.9 & -11.6 & 7.4 & 26.3 & 4.4 \\
1176.5 & 35.1 & 6.9 & 34.0 & 4.4 \\
1199.5 & 17.0 & 4.4 & 33.9 & 3.5 \\
1222.5 & -11.7 & 7.4 & 23.6 & 4.0 \\
1245.8 & 0.9 & 4.2 & 36.2 & 3.7 \\
1269.7 & 7.2 & 3.2 & 73.3 & 5.7 \\
1297.1 & 34.2 & 6.1 & 39.8 & 4.7 \\
1323.5 & 46.6 & 9.9 & 24.5 & 4.1 \\
1346.4 & -2.6 & 4.2 & 51.1 & 5.2 \\
1370.3 & 4.9 & 8.7 & 21.8 & 4.4 \\
1394.3 & -2.6 & 3.2 & 46.6 & 3.6 \\
1420.1 & 12.2 & 5.1 & 32.1 & 3.7 \\
1445.8 & 0.9 & 4.5 & 35.5 & 3.6 \\
1474.9 & 11.6 & 4.9 & 37.7 & 4.1 \\
1503.4 & -8.2 & 5.9 & 39.7 & 5.7 \\
1526.6 & 3.0 & 5.5 & 44.6 & 5.9 \\
1550.4 & 35.5 & 7.8 & 31.3 & 4.5 \\
1576.1 & -4.7 & 4.4 & 33.5 & 3.4 \\
\enddata
 \tablecomments{Phase offsets ($\phi$) are measured with respect to the secondary eclipse.}
\end{deluxetable}





\subsection{Comparison with Armstrong et al.} 
\label{sec:comparison with armstrong et al.}



\begin{figure*}[ht!]
\centering
	\includegraphics[width=6.5in]{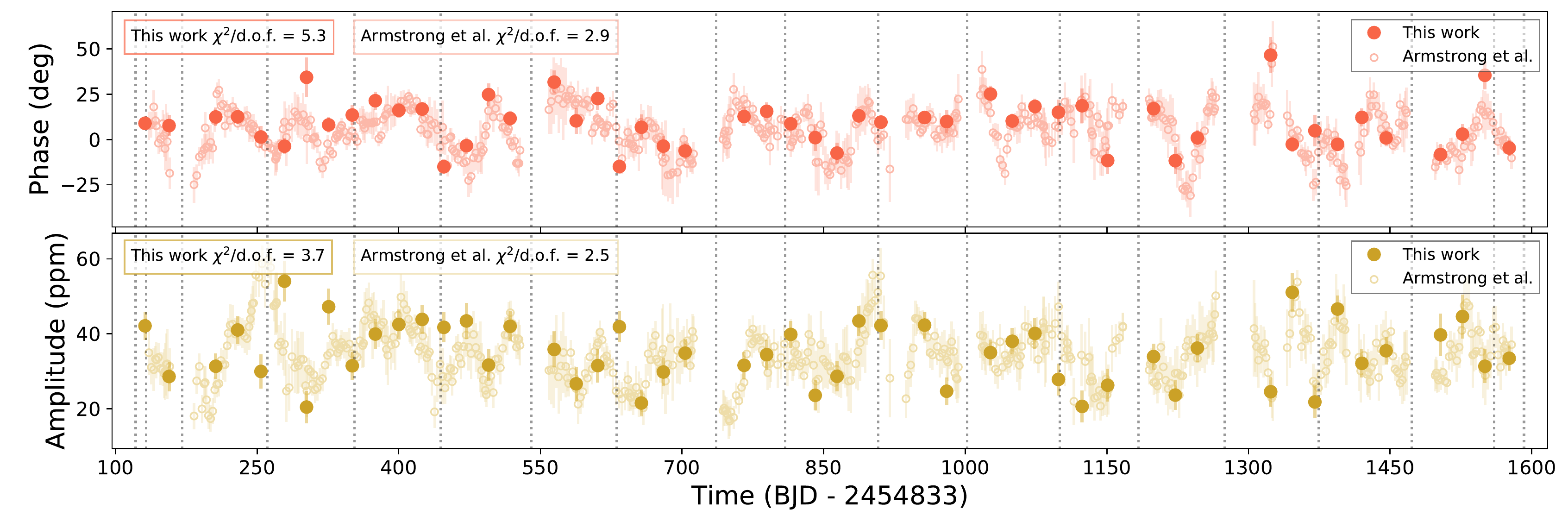}
    \caption{Measurements of the phase (\textit{Top}) and amplitude (\textit{Bottom}) of \thisplanet's phase curve over time. Our measurements are shown in bold points, plotted with the phase offsets and amplitudes from \citet{armstrong2016} shown in open points. The vertical dotted lines show where the \Kepler\ light curve is broken into quarters. The $\chi^2$ per degree of freedom value is shown for each result. Though our result appears to show significant atmospheric variability, the identified variations may be due to unrelated factors.}
    \label{fig:60_and_arm}
\end{figure*}

We compared our apparently statistically significant detection of variability in the phase offset and amplitude of \thisplanet's phase curve with the previous detection of \citet{armstrong2016}. Figure \ref{fig:60_and_arm} shows our phase and amplitude measurements compared to those reported by \citet{armstrong2016}. \bmaroon{In this figure, we discarded outlier points from our results which lie close to the \Kepler\ quarter breaks, in sections of data which were also discarded by \citet{armstrong2016}}. In general, we find that our measurements tend to match those of \citet{armstrong2016}, although their measurements yield a lower reduced $\chi^2/d.o.f.$ value because they used a different method to estimate their uncertainties and report slightly larger values.

Despite the similar values of phase and amplitude measurements, there is an immediately obvious difference between the presentation of our results and those of \citet{armstrong2016} due to the way the phase curve variations were sampled. While both we and \citet{armstrong2016} broke the light curve into segments with a length of 10 planet orbits, we report results for 60 fully independent segments (each including about 10 distinct planet orbits) with only one phase and amplitude measurement for each of those segments - that is, each \Kepler\ flux measurement was only used in one particular MCMC fit. On the other hand, \citet{armstrong2016} oversampled their measured phase variations by a factor of 10 by fitting light curve segments with a length of 10 planet orbits, shifting by only one planet orbit at a time - that is, \citet{armstrong2016} used each \Kepler\ flux measurement in ten different MCMC fits. 

\begin{figure*}[ht!]
\centering
	\includegraphics[width=5.5in]{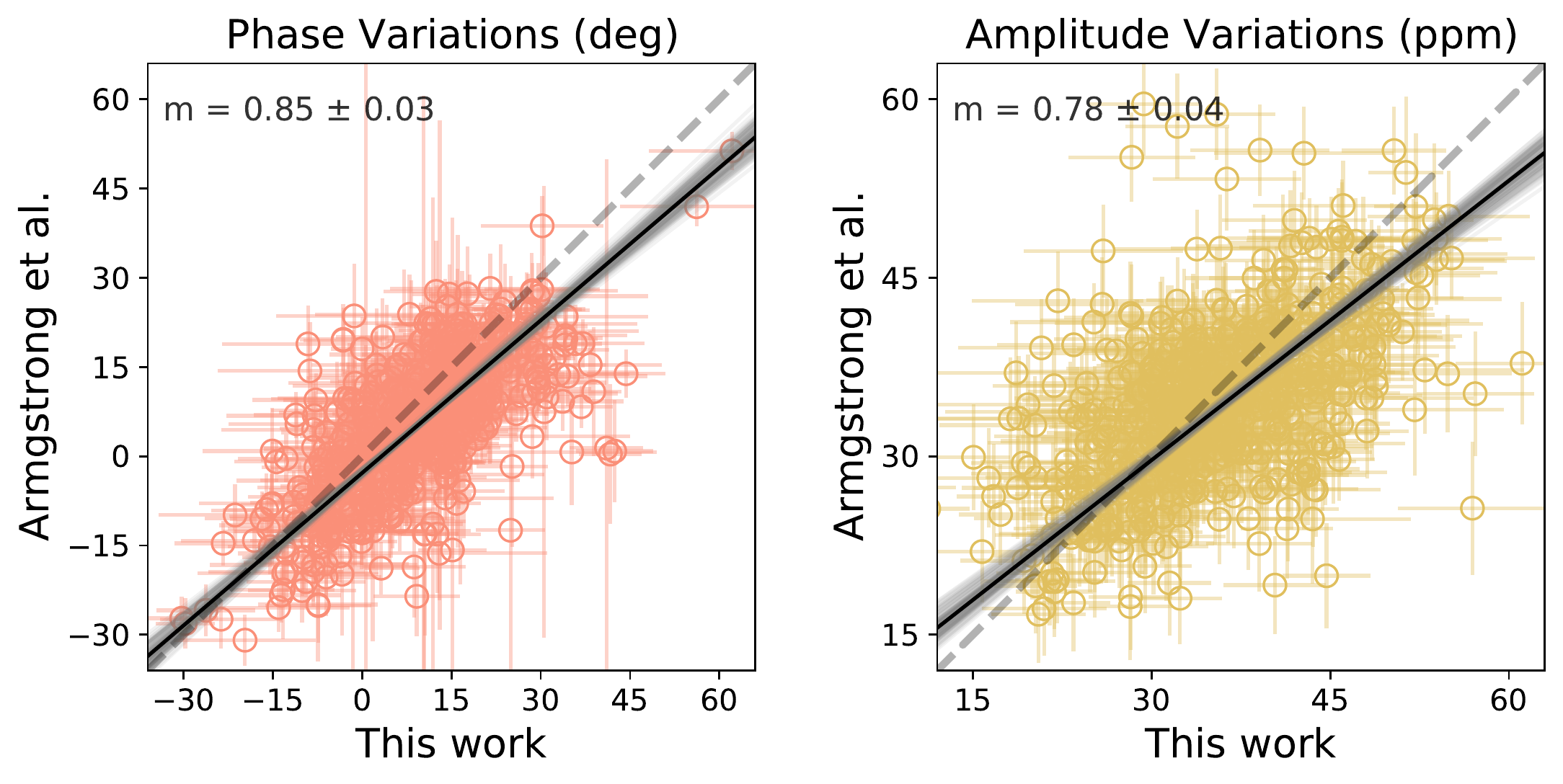}
    \caption{Comparison between our measurements of the phase (\textit{Left}) and amplitude (\textit{Right}) of \thisplanet's phase curve with those from \citet{armstrong2016}. \bmaroon{The dashed gray diagonal lines in each plot show a one-to-one slope, while the solid black line shows the slope from a fit accommodating the error bars in both axes. The fitted slope and uncertainty appears on the top left of each plot.} The phases and amplitudes we measure in each light curve segments are fairly similar to those measured by \citet{armstrong2016}.}
    \label{fig:me_vs_arm}
\end{figure*}

In order to compare our results to those of Armstrong et al., we replicated their sampling by sliding a 10-orbit window across the light curve and oversampling our phase curve measurements by a factor of 10. We then interpolated our oversampled results to match the times from \citet{armstrong2016} and compare the results in Figure \ref{fig:me_vs_arm}.  While some of the points differ, we were generally able to reproduce their phase and amplitude measurements results. The differences between our measurements are likely due to slight differences in our analysis, like the details of the flattening procedure and the fact that we excluded the planet's secondary eclipses from our modeling, while \citet{armstrong2016} included them.

\subsection{How Sliding Windows can give the appearance of real time correlations}
\label{sec:how sliding windows can give the appearance of real signals}

The main difference between the presentation of our measurements of the phase and amplitude of \thisplanet's phase curve and those made by \citet{armstrong2016} is the fact that they presented oversampled measurements from partially overlapping light curve segments. This sampling/data visualization strategy can be helpful in identifying subtle features in time series observations, but it can also suggest the presence of coherent variations when none are present. In this section, we illustrate how this sampling strategy can give the appearance of correlated variability even when the source dataset contains no such correlations. 

\begin{figure*}[ht!]
\centering
	\includegraphics[width=6.5in]{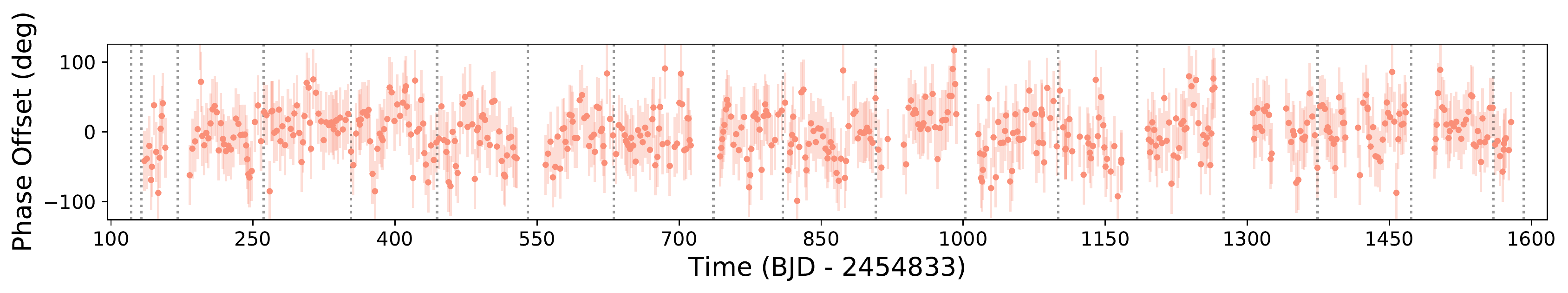}
	\includegraphics[width=6.5in]{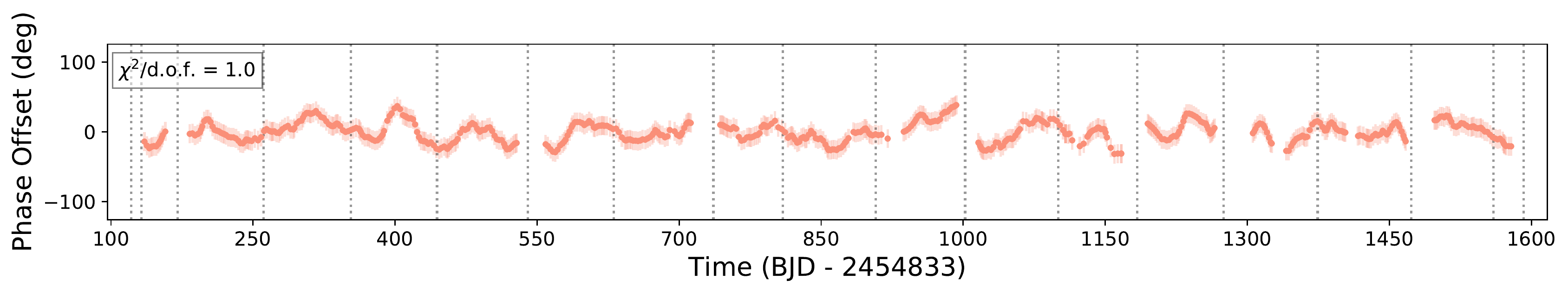}
	\includegraphics[width=6.5in]{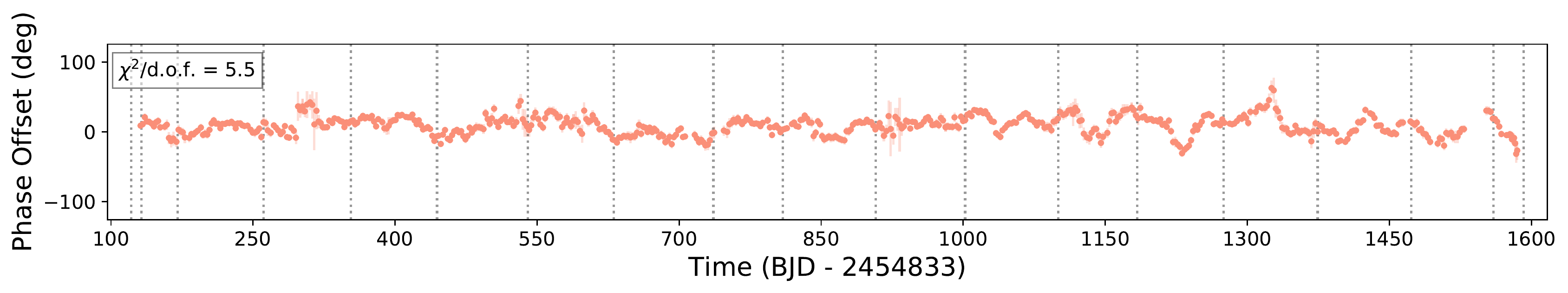}
    \caption{An illustration of how oversampling can create the appearance of variability in time series. \textit{Top:} A time series of Gaussian random noise, sampled at the times of the measurements of \thisplanet's orbital phase by \citet{armstrong2016}. \textit{Middle:} The same distribution of Gaussian random noise shown in the top panel, but smoothed by a box filter with an oversampling factor of 10. \textit{Bottom:} Our phase offset variation result, sampled using a sliding window of 10 orbits. In all three panels, the vertical dotted lines show where the \Kepler\ light curve is broken into quarters. The sliding window makes the result appear smoother by oversampling the data, but qualitatively similar to the smooth variations in the Gaussian random noise in the previous panel.} 
    \label{fig:random_noise}
\end{figure*}

We show an example of this phenomenon in Figure \ref{fig:random_noise}. This figure shows a time series of purely Gaussian random numbers, sampled at the times of measurements of \thisplanet's phase curve by \citet{armstrong2016}. In the first panel, the Gaussian random numbers are shown as if they were phase measurements from individual orbits of \thisplanet, while in the second panel, the random numbers have been smoothed by a 10 point boxcar filter, simulating the effect of averaging 10 orbits together while shifting the window by only one orbit at a time. Clear time correlations are visible in the smoothed plot, even though the underlying data is purely white noise. This correlation structure is introduced because each averaged point includes 90\% of the same data as its neighboring points. This is an expected result because smoothing with a boxcar window is equivalent to suppressing high-frequencies in Fourier space, leaving only slow variations with the appearance of coherent changes. 

Finally, we show our measurements of the phase of \thisplanet's phase curve in the bottom panel of Figure \ref{fig:random_noise}, when we oversample the data by of a factor of 10, following \citet{armstrong2016}. Although the uncertainties on the individual points are smaller than the uncertainties on the smoothed Gaussian random noise time series, the time correlations visible in the oversampled time series are qualitatively similar to those in the smoothed Gaussian random noise time series, indicating that we cannot rely on the appearance of this time series to confirm that the apparent variations in \thisplanet's phase curve are astrophysical.

\subsection{Assessing evidence for astrophysical variability}
\label{sec:assessing evidence for astrophysical variability}
\begin{figure*}[hp!]
\centering
    \raggedleft
	\includegraphics[width=7in]{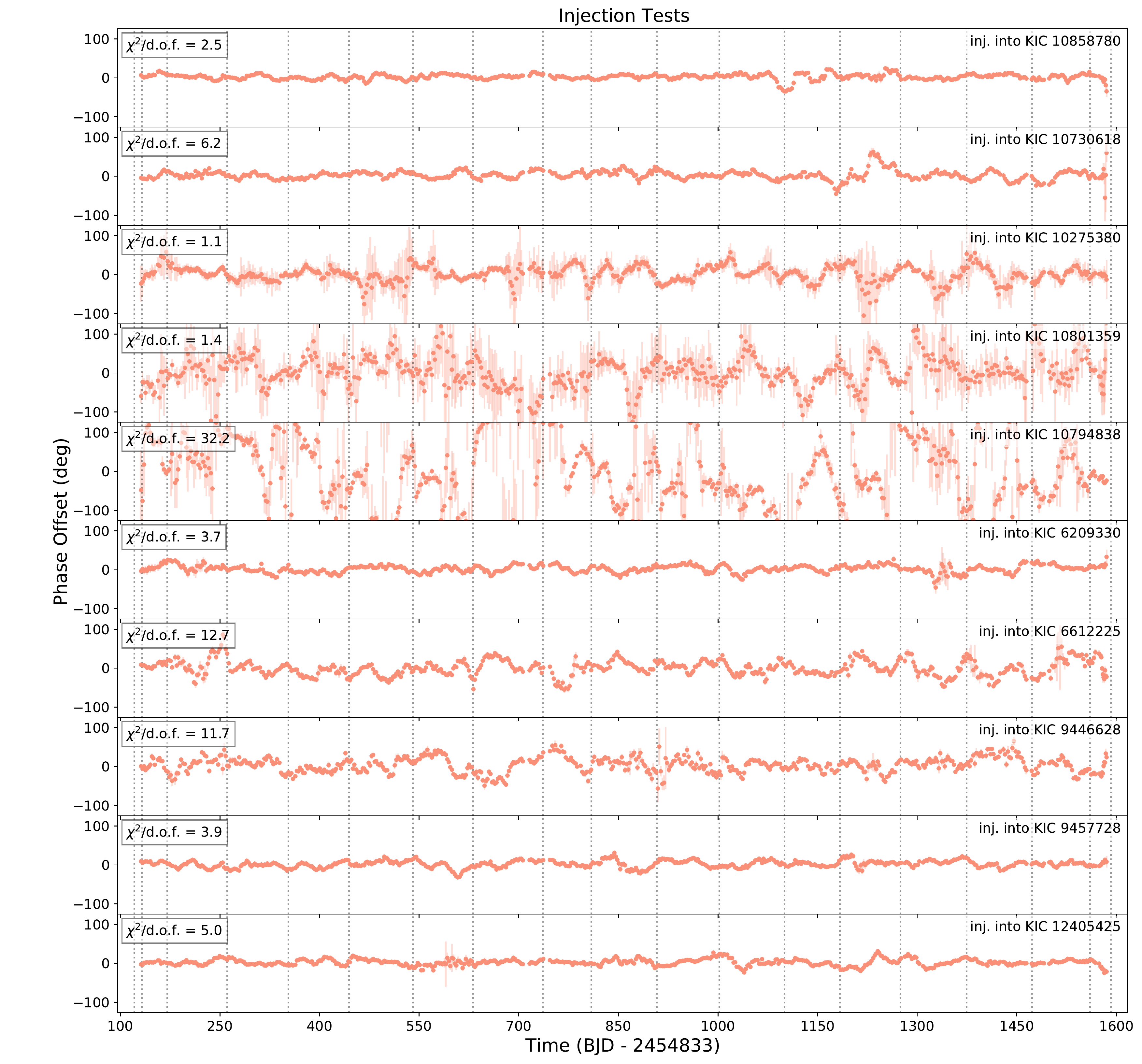}
	\includegraphics[width=6.74in]{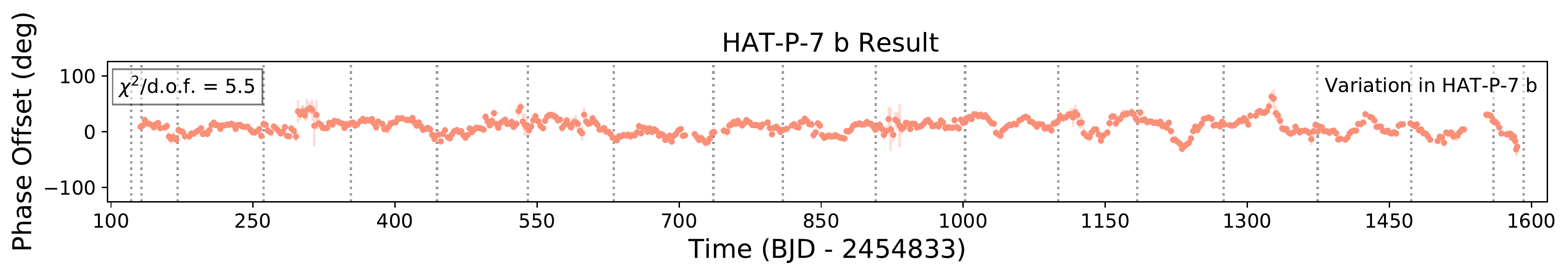}
    \caption{Measurements of the phase offset of the peak of the injected phase curves for each of the different stars in our injection/recovery tests. The \Kepler\ Input Catalog (KIC) identifier of the injected star is shown in each panel, along with the reduced $\chi^2/d.o.f.$ value for each time series. The vertical dotted lines show where the \Kepler\ light curves are broken into quarters. Even though we know the true signal injected into these planets was stationary and did not vary in time, we measure strong variations in these time series. Qualitatively, these variations are similar to those found in the actual light curve of \thisplanet, and quantitatively, their apparent statistical significance can exceed that of the variations in \thisplanet's phase curve. Our oversampled phase offset variation result for \thisplanet\ is shown at the bottom for comparison. It is possible that the variations measured in \thisplanet's phase curve could be caused by the same effects as in our injection/recovery tests.}
    \label{fig:all_inj_phase}
\end{figure*}

We have measured the phase and amplitude of \thisplanet's phase curve and found that our measured phases and amplitudes are in good agreement with those of \citet{armstrong2016}. However, it is not clear whether the apparently statistically significant variations we measure are actually due to astrophysical variability in \thisplanet's atmosphere, or some other effect. We therefore test whether it is possible that the detected variations could be caused by factors other than true variability in \thisplanet's atmosphere. 

We do this using the results of our injection/recovery tests described in Section \ref{sec:injection/recovery}. In these tests, we injected non-varying phase curve signals into \Kepler\ light curves of stars similar to \thisstar. We then repeated our analysis methods (described in Section \ref{baseline}) on these injected signals to see if we would find incidental evidence of variability where there was no astrophysical variability. We conducted the injection/recovery tests using 10x oversampling strategy used by \citet{armstrong2016} and discussed in Section \ref{sec:how sliding windows can give the appearance of real signals}. 

Figures \ref{fig:all_inj_phase} and \ref{fig:all_inj_amp} show the recovered phases and amplitudes of the injected non-variable phase curves, as well as the $\chi^2/d.o.f.$ value (assuming a constant phase and amplitude in time) for each injection result. The oversampled time series of the recovered phase for these non-varying phase curves are qualitatively similar to the recovered phases for \thisplanet, and their $\chi^2/d.o.f.$ values are similar, and in some cases exceed, those of the apparent variability in \thisplanet. \bmaroon{In general, we observe that injected stars with easily visible, high-amplitude photometric variability yield the largest $\chi^2/d.o.f.$, but even apparently quiet stars show $\chi^2/d.o.f.$ values comparable to \thisstar.} Although the variations in \thisplanet's atmosphere appear statistically significant at first, a comparison to the injected results show that similarly strong apparent variations can be caused by factors other than a varying exoplanet atmosphere. 

We note that \citet{armstrong2016} also performed injection tests (albeit on a smaller number of stars) and concluded that the variations they measured in the amplitude of \thisplanet's phase curve could be spurious, while they considered the phase \bmaroon{offset} variations robust. We agree with their conclusion that the variations in \thisplanet's phase curve amplitude do not appear to be robust, but our analysis with a larger set of stars suggests that the variations in the phase \bmaroon{offset} may not be robust either. 

\begin{figure}[ht!]
\centering
	\includegraphics[width=3.5in]{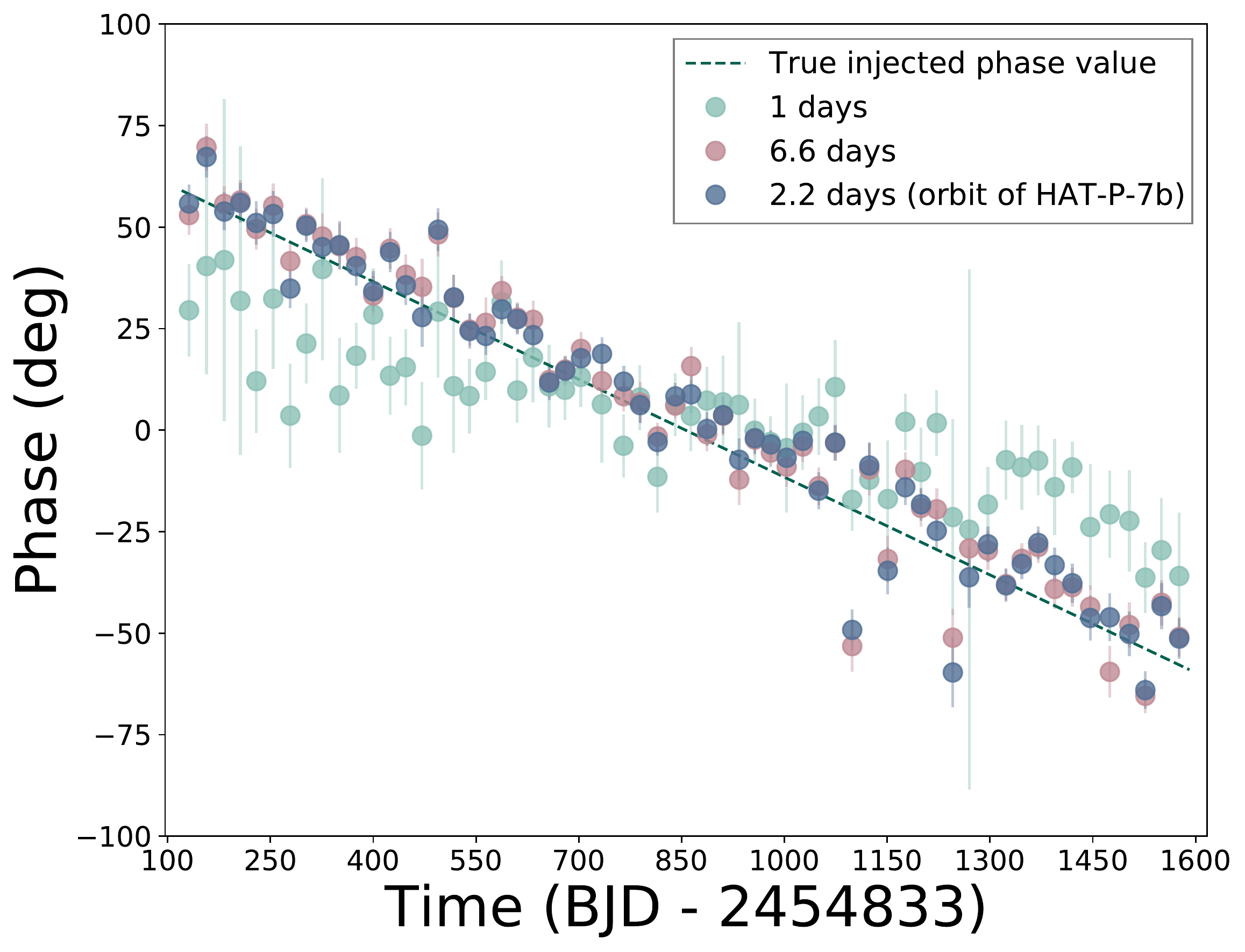}
    \caption{\bmaroon{Measurements of the phase offset recovered from the process described in Section \ref{sec:assessing evidence for astrophysical variability}. The average phase curve of \thisplanet\ was divided out of the Kepler light curve, and instead, a phase curve shift varying from $\pm 60 \deg$ was injected. The plot shows several recovered phase offsets from the same injected light curve, but processed with different spline spacings. The dotted line shows the true value of the injected phase offset. As expected from our previous tests and shown in Figure \ref{fig:brk}, spline spacings shorter than the period of \thisplanet\ do attenuate real variations; however, with a basis spline spaced at the period of \thisplanet\ or longer, our analysis does not flatten away real phase variations.}}
    \label{fig:rev}
\end{figure}

\bmaroon{In order to ensure that our analysis methods do not attenuate real variations, we repeated the injection/recovery tests, but instead injected a varying phase signal. The average phase curve over the entire Kepler dataset of \thisplanet\ was divided out of the light curve in order to remove the real effects of \thisplanet. A phase curve with a phase offset shifting from $+60\deg$ to $-60\deg$ over the course of the \Kepler\ observations was injected back into the flattened \thisstar\ light curve. We attempted to recover the injected shifting phase using our analysis methods at various spline spacings. These results are shown in Figure \ref{fig:rev}. While spline spacings shorter than the period of \thisplanet\ do attenuate the recovered phase offsets, analysis using spline spacings of the period of \thisplanet\ or longer preserve real phase offset variations.}

\begin{figure*}[hp!]
\centering
    \raggedleft
	\includegraphics[width=7in]{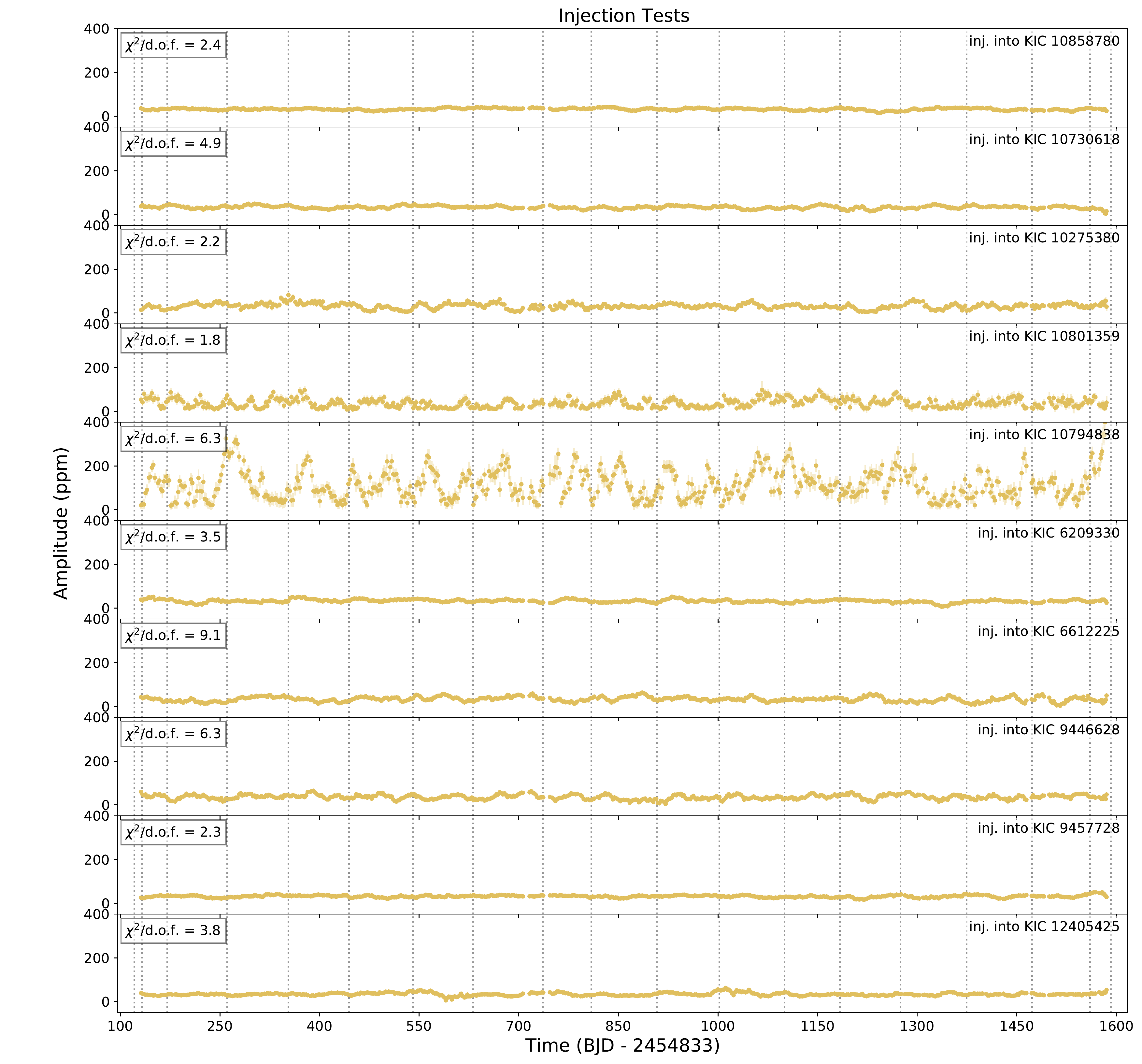}
	\includegraphics[width=6.67in]{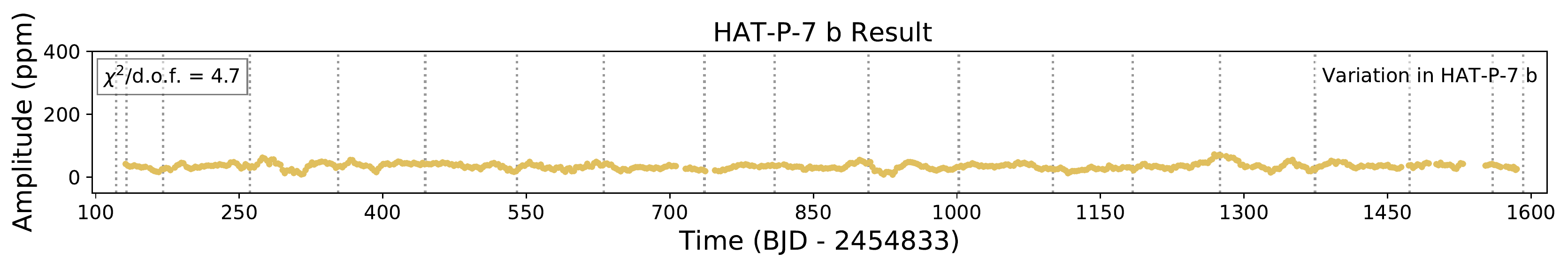}
    \caption{Measurements of the amplitude of the injected phase curves for each of the different stars in our injection/recovery tests. Like Figure \ref{fig:all_inj_phase}, the KIC identifier of the injected star is shown in each panel, along with the reduced $\chi^2/d.o.f.$ value for each time series. The vertical dotted lines show where the \Kepler\ light curves are broken into quarters. The amplitude time series also show apparent variations, even though the injected signal had no time variations. There are apparently effects other than true atmospheric variability that can introduce variations as large as we see in \thisplanet's phase curve. Our oversampled amplitude variation result for \thisplanet\ is shown at the bottom for comparison.\bmaroon{ We note that while the stars with the highest $\chi^2/d.o.f.$ values show high-amplitude photometric variability, even apparently quiet stars yield $\chi^2/d.o.f.$ values similar to \thisstar.} }
    \label{fig:all_inj_amp}
\end{figure*}





\section{\bmaroon{Possible causes of spurious variability signals}} \label{causes}
\bmaroon{Since our injection/recovery tests showed that phase offset variations like those we observe in \thisplanet's phase curve be spurious,} we investigated what other factors besides true variations in planet atmospheres could cause the variations we measure in our injected phase curves, and potentially in \thisplanet\ as well.  Based on our tests of the robustness of our analysis, the choice that affected our measured phase curve parameters the most was the way we flattened the light curve \bmaroon{to remove} low-frequency stellar/instrumental variability. Given the sensitivity of our results to changes in how low-frequency variability was removed, we hypothesized that residual unfiltered low-frequency variability could be causing the apparently spurious changes to the phase curve parameters.

\subsection{\bmaroon{Low-frequency variability in injected light curves}}
\begin{figure}[ht!]
    \centering
    	\includegraphics[width=3.5in]{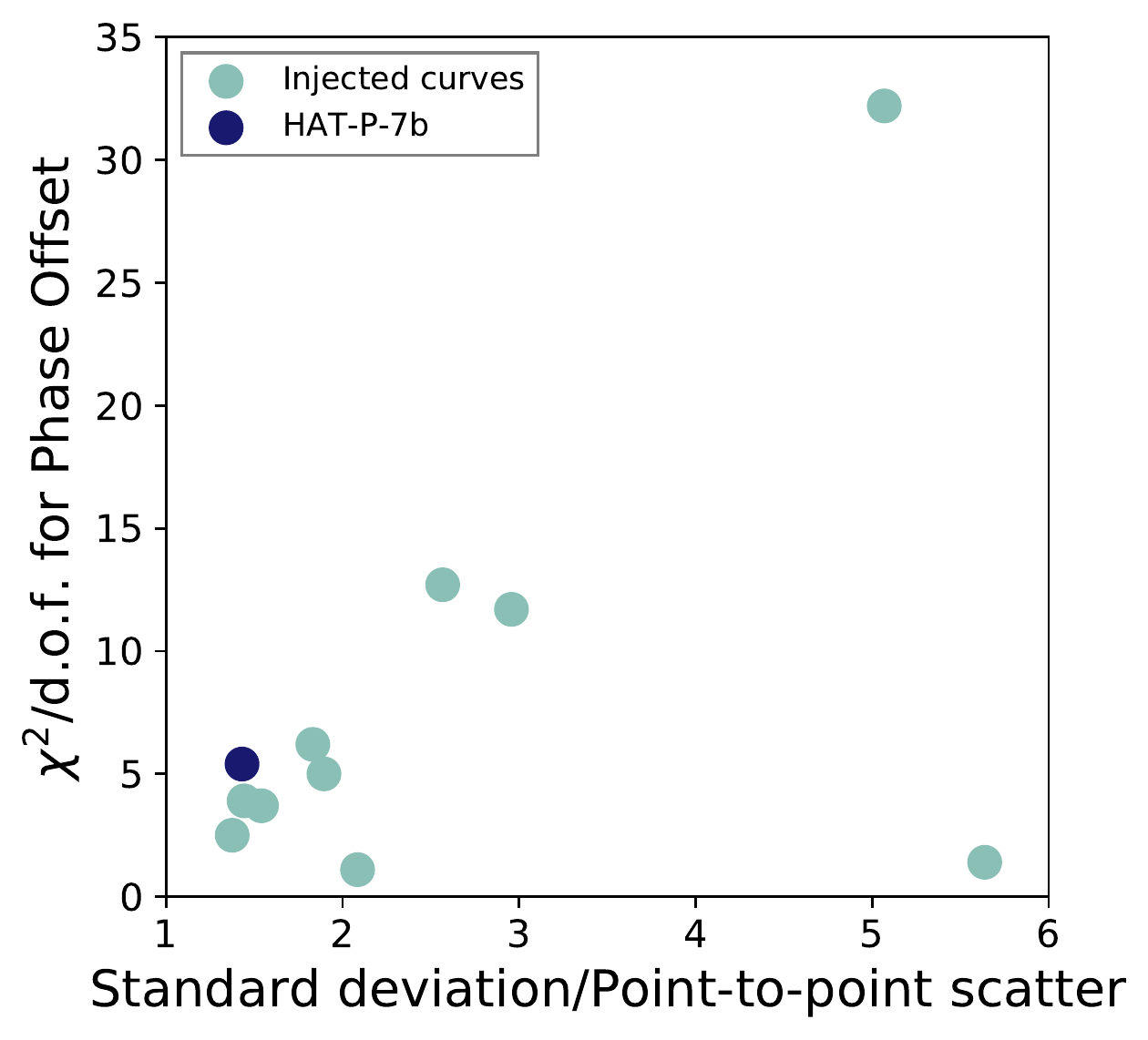}
        \caption{A possible relationship between the level of apparent variations in the phase curve parameters and the amount of low-frequency variability in injected light curves.  The x-axis shows the ratio between the standard deviation of the \Kepler\ light curve (excluding transits and eclipses) and the point-to-point scatter of the same. A larger ratio indicates dominance of longer-scale variations, such as those from stellar rotation signals. The y-axis shows the $\chi^2/d.o.f.$ value of measurements the phase ($\phi$) of the sine at the planet's orbital period over time. Injected curves are shown in green, while \thisstar\ is shown in dark blue. The possible relationship between this metric designed to quantify red noise in the light curve and the apparent statistical significance (using $\chi^2/d.o.f.$ as a proxy) of variations implies that the variations seen in \thisplanet's phase curve may be due to these low-frequency variations. KIC10801359 is the outlier visible in the lower right-hand corner of the plot. It shows $\sim50-500$ ppm amplitude photometric variability on short ($\sim 1$ day) timescales, but is qualitatively similar to other noisy light curves that show higher reduced $chi^2/d.o.f.$ values.}
        \label{fig:stddev_p2p}
\end{figure}

\bmaroon{As a first test of this theory}, we quantified the low-frequency variability in each of the injected light curves by calculating the ratio of the overall scatter in the light curve to scatter on short timescales. In particular, we calculated the standard deviation of the light curve divided by the point to point scatter (e.g. the P2P metric discussed in Section 4.2 of \citealt{aigrain2014}). When this ratio is near unity, the light curve is dominated by variations on timescales of one \Kepler\ long-cadence exposure (30 minutes), but when this ratio is large, the light curve is dominated by slow variability.  Figure \ref{fig:stddev_p2p} shows the reduced $\chi^2/d.o.f.$ in the phase offset for each of the stars from our injection/recovery tests plotted against the ratio of slow to fast variations (standard deviation/point to point scatter) in the light curve. We find a tentative trend  that appears to show that $\chi^2/d.o.f.$ values are highest for stars for which slow variations dominate. \bmaroon{This indicates that the recovered variations in our injected phase curves could be due to residual slow variations in the star's light curve, and that this red noise could contribute to the variations detected in \thisplanet\ as well.}

\subsection{\bmaroon{Excess noise}}

\bmaroon{To explore further the possibility that unfiltered stellar photometric variations contributed to the apparent variability we observe in \thisplanet's phase curve result, we investigated how the goodness-of-fit of an individual phase curve segment correlates with the measured phase curve offset.} 
\bmaroon{To do this, we quantified the amount of excess photometric scatter within an individual phase curve segment. We defined an ``excess noise'' term,  $\sigma_{excess}$, as follows:} 
\begin{equation}\label{excess_noise}
        \sigma_{excess} = \sqrt{\sigma_{K}^2 - \sigma_{lc}^2}
\end{equation}
\bmaroon{where $\sigma_{K}$ is the uncertainty recovered by our MCMC fitting, and $\sigma_{lc}$ is the median uncertainty of the star's Kepler light curve flux data. The excess noise term,  $\sigma_{excess}$, quantifies how much noisier a given light curve segment is compared to the expectation from the \Kepler\ instrumental uncertainties, assuming the noise sources are independent and Gaussian. Because \Kepler\ was an exceptionally well-behaved instrument, we expect that the excess noise we measure is dominated by variability from the stars themselves.}

\begin{figure*}[ht!]
    \centering
    	\includegraphics[width=5.5in]{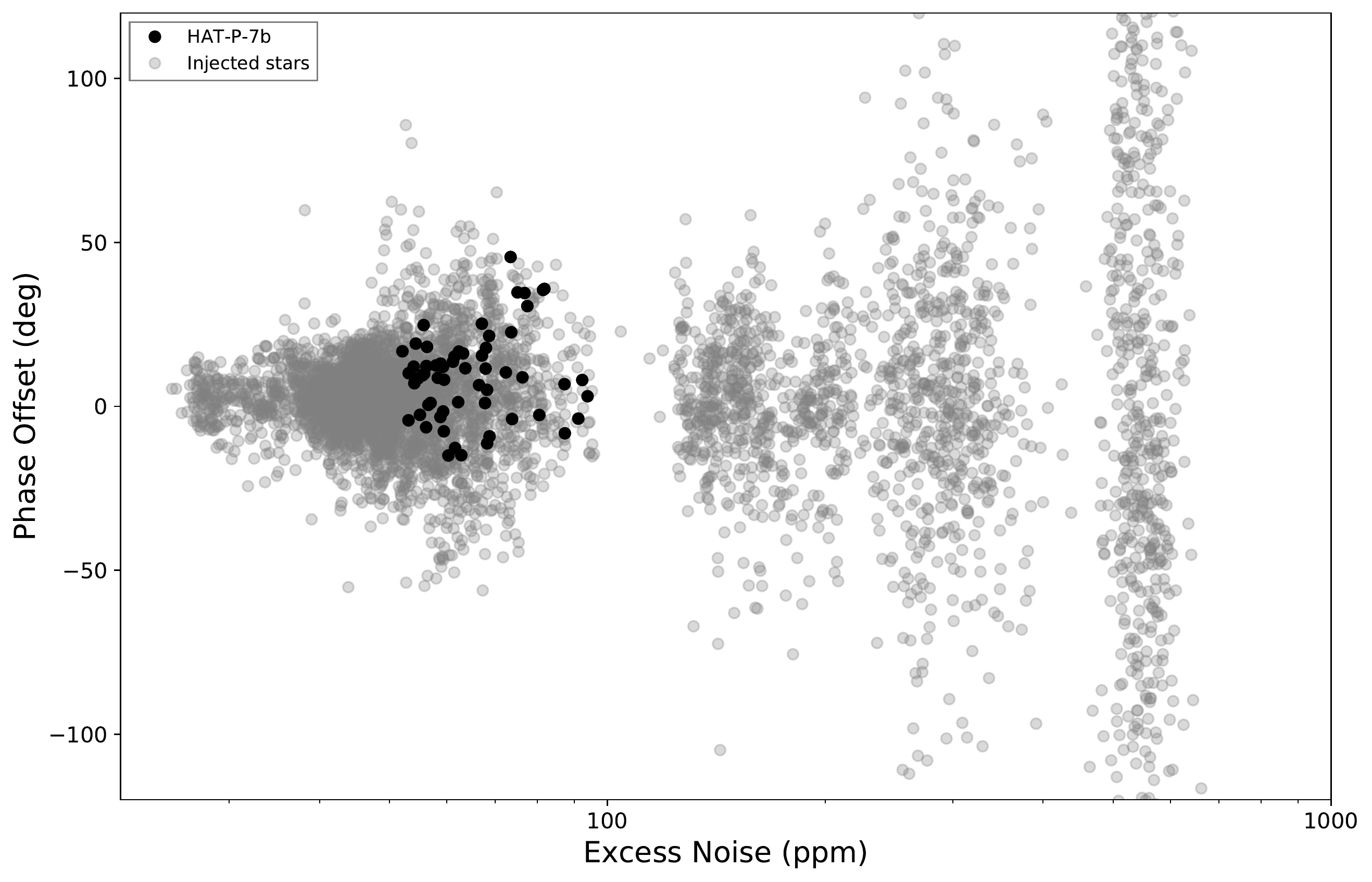}
    	\includegraphics[width=5.5in]{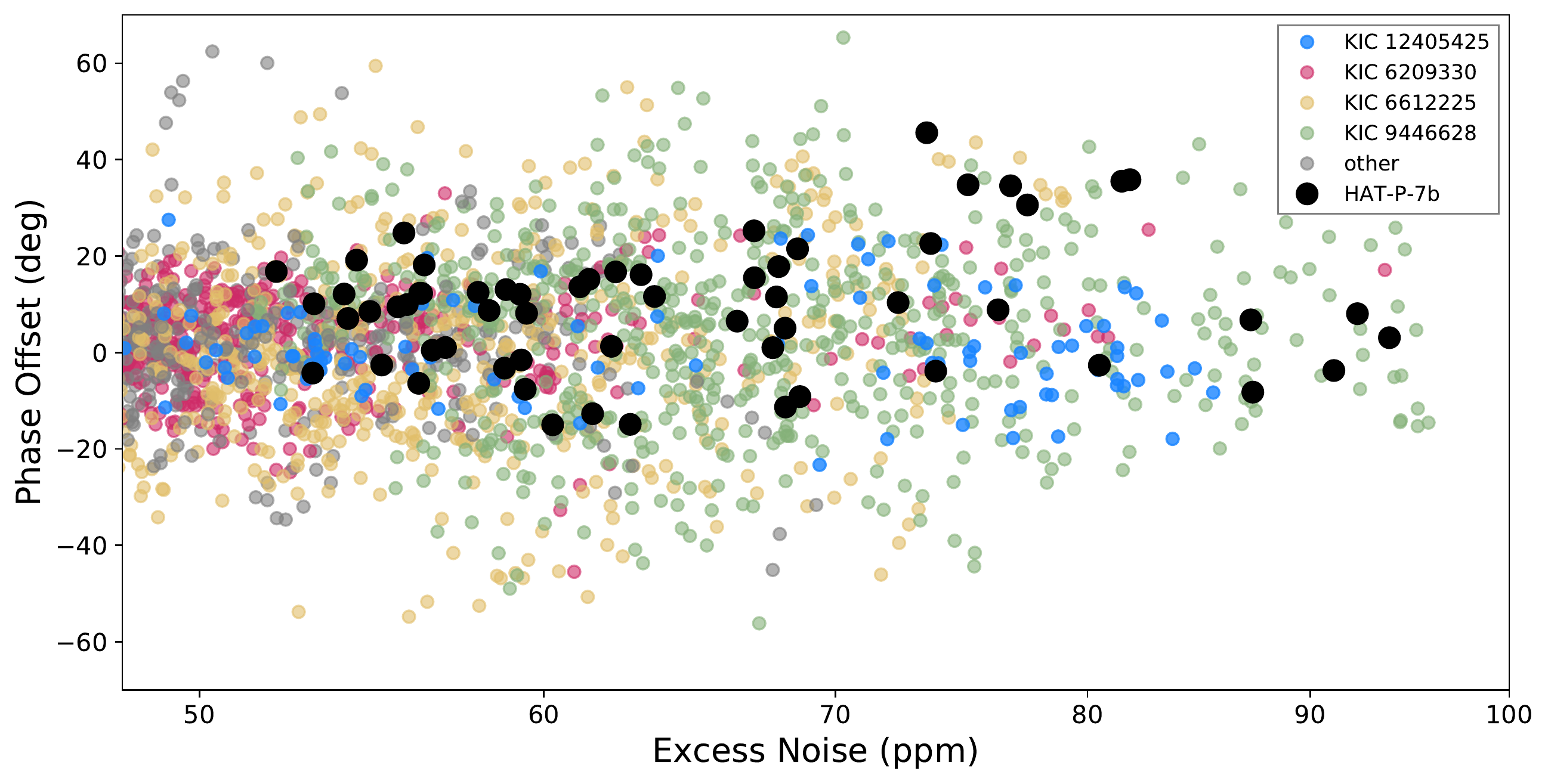}
        \caption{\bmaroon{Excess noise vs recovered phase offset in \thisplanet\ and the injected light curves. Excess noise is defined in Equation \ref{excess_noise}, where a high excess noise indicates a poor fit quality. The high excess noise of some of the injected light curves (KIC10801359, KIC10794838) makes them poor comparisons against the \thisplanet\ result. However, some of the injections which show comparable excess noise to the \thisplanet\ result also show a similar phase offset range (the injections in the same range as \thisplanet\ are shown in color in the second plot). This suggests that stellar variability can cause similar phase offset variability results as those we recovered for \thisplanet.}}
        \label{fig:excess_noise}
\end{figure*}

\bmaroon{After calculating  $\sigma_{excess}$ for each phase curve segment for \thisstar\ and all of our injected light curves, we searched for a relationship between the excess noise within each phase curve segment and the measured phase offset (see Figure \ref{fig:excess_noise}). As we would expect, segments with larger values of $\sigma_{excess}$ show larger phase offset variations for injected light curves. Interestingly, the phase offset variations and excess noise levels we measured in \thisplanet's phase curve fit in well with the trend from the injection/recovery tests. The bottom panel of Figure \ref{fig:excess_noise} shows that several different stars from our injection/recovery tests have both similar phase curve variations and excess noise levels as \thisstar. The fact that \thisplanet's apparent phase curve variations are similar to those found in stars from our injection/recovery tests shows that excess noise (likely due to stellar variability) could contribute to \thisplanet's apparent phase offset variability.}


\subsection{Astrophysical noise sources in \thisstar's light curve}
\label{sec:astrophysical noise sources in hatp7b's light curve}

\bmaroon{Because our tests suggested that excess noise in \thisstar's light curve could be driving at least some of the variations we see in \thisplanet's phase curve parameters, we investigated the sources of noise in \thisstar's light curve. We calculated a Lomb-Scargle periodogram (for more information about the Lomb-Scargle algorithm, see \citealt{vanderplas_lombscargle}) of the full PDC corrected short-cadence light curve of \thisstar. We first removed the transits, secondary eclipses, and phase curve of \thisplanet, by dividing the light curve by the average model of \thisplanet's signal that we calculated in Section \ref{sec:injection/recovery} for our injection/recovery tests. We also pre-processed the light curve by fitting a basis spline with knots spaced every 6 days to the light curve, and dividing the light curve by this spline curve. This removed all variations on timescales longer than about 6 days. After calculating the Lomb-Scargle periodogram, we converted from the default output in units of power spectral density ($\mathcal{D}$) to units of power ($p$) and amplitude ($a$), using the method described in Appendix A of \citet{kjeldsen_lombscargle}. We tested the calibration of this conversion by creating an artificial signal, taking its transform, and confirming that the periodogram amplitude units properly corresponded to that artificial input signal. We found the conversion to be: 
\begin{equation}\label{PDS_to_amp}
        a = \sqrt{p} = \sqrt{\frac{4\mathcal{D}}{N}}
\end{equation}
where $N$ is the number of measurements in the time series.}

\begin{figure*}[t!]
    \centering
    	\includegraphics[width=\textwidth]{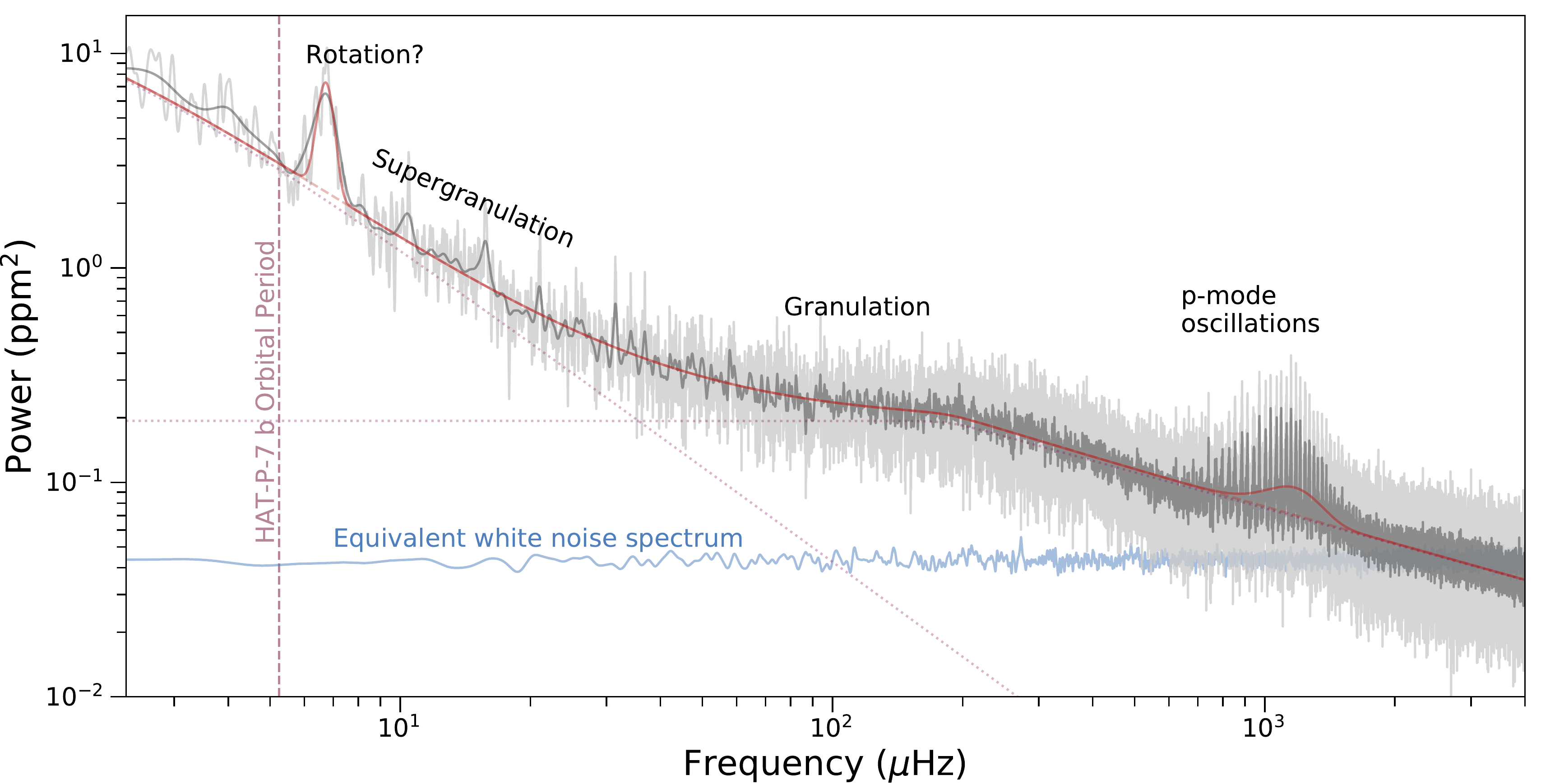}
        \caption{\bmaroon{The power spectrum of \thisstar's light curve, after removal of \thisplanet's signal. The light and dark grey curves show \thisstar's power spectrum smoothed by convolution with a Gaussian envelope with standard deviations of 0.3 \microhertz\ and 2.4 \microhertz, respectively. The red curve shows the best-fit model to the power spectrum, and dashed lines show individual components of the fit. \thisstar's power spectrum shows evidence for several different signals, at least two of which (namely the broad supergranulation signal and the possible rotation signal at 1.7 days or 6.7 \microhertz) have timescales similar to \thisplanet's orbital period. The blue curve is the power spectrum of an array of Gaussian random white noise with the same time axis and standard deviation as \thisstar's light curve (with the planetary signal removed). Compared to the power spectrum of purely white noise, \thisstar's light curve shows excess noise on timescales close to the planetary orbital period that is unaccounted for in our phase curve analysis. The excess correlated astrophysical noise in \thisstar's light curve could contribute to the apparent variations in \thisplanet's phase curve.}}
        \label{fig:fullpowerspectrum}
\end{figure*}

\bmaroon{We show \thisstar's power spectrum in Figure \ref{fig:fullpowerspectrum}. The power spectrum shows evidence for several different astrophysical processes:}
\begin{enumerate}
\item \bmaroon{Solar-like p-mode oscillations which appear at high frequencies ($\approx 1100 \microhertz$ or 15 minute periods). P-mode oscillations manifest as a forest of peaks in the power spectrum with a roughly Gaussian envelope on timescales of 5 minutes for sun-like dwarf stars and provide a wealth of information about the detailed properties of the star, including highly precise mass, radius, age, and inclination estimates (for a thorough review of the subject, see \citealt{chaplin}). The p-mode oscillation spectrum of \thisstar, which has been studied extensively by previous authors \citep{jcd, lund, benomar, campante2016}, peaks at lower frequencies than the Sun because \thisstar\ is a slightly evolved subgiant. Nevertheless, the timescale of these variations is so short that it should not affect our measurement of \thisplanet's phase curve on timescales two orders of magnitude longer. }

\item \bmaroon{Granulation, which manifests as a flat noise source at low frequencies, up to a break around 200 \microhertz\ or 1.4 hours, after which it falls off as a power law. Photometric granulation signals are due to changes in brightness as warm convective bubbles arrive at the surface of the star forming bright spots, while cooler (and less bright) gas falls downwards below the photosphere. This process introduces stochastic changes in brightness that have been observed in the Sun \citep{domingo, aigrain2004} and numerous other stars \citep{kallinger}. However, the timescales where granulation dominates are much shorter than \thisplanet's orbital period, and will not affect our measurement of it. }

\item \bmaroon{A similar process called supergranulation manifests in \thisstar's power spectrum as another power law in frequency space that dominates at frequencies lower than about 20 \microhertz\ or timescales longer than about 14 hours. Supergranulation is a fluid-dynamical phenomenon similar to granulation, but which takes places on longer timescales, primarily involves horizontal flows, and is less-well understood than its shorter-timescale cousin \citep{supergranulation}. Supergranulation has been detected in the power spectrum of the Sun \citep{harvey1985} and other stars \citep[e.g.][]{bazot}, but has not been as well studied as granulation. Because the supergranulation signal exhibits larger photometric amplitudes on longer timescales than granulation, and indeed overlaps the orbital period of \thisplanet, it could plausibly affect our measurement of the planetary phase curve parameters. }

\item \bmaroon{A well localized peak at about 1.7 days, which we cannot positively identify, could be related to the stellar rotation period. Periodic (or quasi-periodic) signals at the stellar rotation period are commonly found in \Kepler\ light curves due to surface inhomogeneities (like starspots) rotating in and out of view and manifest with similar properties to the 1.7 day signal we see in \thisstar's light curve. If the 1.7 day periodicity we see is indeed caused by stellar rotation, it would require the star's rotational axis to be nearly pole on to yield low projected rotation velocity \citep{lund}. While an analysis of photometric anomalies due to gravity darkening in \thisplanet's transit light curve are consistent with a 1.7 day rotation period \citep{masuda}, it seems to be inconsistent with asteroseismic measurements of the stellar inclination \citep{lund}. While we cannot conclusively determine the source of this signal, we note that since its timescale is close to that of the planetary orbital period and its amplitude is greater than any other astrophysical signal in the light curve, it could plausibly affect our measurement of \thisplanet's phase curve parameters. }
\end{enumerate}

\bmaroon{For illustrative purposes, we modeled \thisstar's power spectrum as a sum of variability from these four processes. Our model was the sum of two Gaussian functions (one to model the envelope of p-mode oscillations around 1100 \microhertz, and one to model the possible rotation signal at 1.7 days), and two ``Harvey-like'' functions to model the broad granulation and super-granulation functions \citep{harvey1985}. In particular, the ``Harvey-like'' functions, $\mathcal{H}$, are given by: }
\begin{equation}
    \mathcal{H} = \frac{\alpha}{(1 + (2\pi\nu\tau)^\gamma)^\beta}
\end{equation}
\noindent \bmaroon{where $\nu$ is the frequency at which the function is evaluated, and $\alpha$, $\beta$, and $\tau$ are free parameters.  We fixed $\gamma$ to 2 for the supergranulation power spectrum and 16 for the granulation power spectrum. Our best-fit model, and the individual ``Harvey-like'' functions are shown in Figure \ref{fig:fullpowerspectrum}. }

\bmaroon{We also tested for the presence of excess correlated noise by generating a light curve with perfectly white noise and comparing its periodogram to that of the actual \thisstar\ light curve. We created an array of normally distributed random noise the same length as the short cadence \Kepler\ light curve of \thisplanet, with a standard deviation equal to that of the short cadence light curve with the planetary signal removed. We then calculated the Lomb-Scargle periodogram of this white noise light curve (using the same time-stamps of the \Kepler\ short cadence light curve)  and normalized it as described above. We then plotted this white noise periodogram in Fig \ref{fig:fullpowerspectrum}. } 


\bmaroon{We found by comparing the periodogram of \thisstar's light curve with that of perfectly white noise that \thisstar's light curve exhibits a significant amount of excess noise on timescales comparable to \thisplanet's orbital period. Because our analysis to measure \thisplanet's phase curve amplitudes and phase offsets did not account for excess noise on these timescales, the presence of such noise could explain the apparently significant variations in the planet's phase curve offset. In Section \ref{sec:linear_fitting_test}, we probe the potential effect of the excess noise due on our phase curve fits. }

\subsection{\bmaroon{Power Spectrum of Short Light Curve Segments}}
\label{sec:source of variation}

\bmaroon{We also calculated the Lomb-Scargle periodogram of the light curve in shorter segments. By inspecting the Lomb-Scargle periodograms of the individual segments, we can be more sensitive to short-timescale variability in the phase curve that would average out over the full light curve. We created the periodogram using the Kepler short-cadence light curve data, with the planetary signal removed by fitting and dividing out transits as well as the median phase curve over the data set (as discussed in Section \ref{sec:assessing evidence for astrophysical variability}). We divided the light curve into 60 segments (as we did for our previous phase curve modeling analysis), and plotted the periodogram for each segment separately. If the phase offset variability signal we detected was in fact due to variations on the planet as opposed to the star, we would expect to see residual phase variations within each shorter segment.}

\bmaroon{Because the planet's rotation period is tidally locked to its orbital period and we observe different longitudes of \thisplanet\ as it rotates every 2.2 days, changes in \thisplanet's atmosphere would show up in a periodogram at its \textit{synodic period}. For slow changes to the atmosphere, as suggested by \citet{armstrong2016}, the synodic period will be close to the planet's orbital period. So, any residual atmospheric variability in the light curve would result in
power spikes near \thisplanet's orbital period.}

\bmaroon{We quantified how close to the planet's orbital period such residual spikes must be in order to suggest an atmospheric variability signal by estimating the range of synodic periods for plausible atmospheric variability signals.  We did this by identifying the fastest phase change from our results in Figure \ref{fig:ourresults}. The largest change in phase offset from one point to the next corresponded to a variability of about $5 \deg$ per day. The range of plausible synodic periods, $P_{\rm syn}$, for atmospheric variability around \thisplanet's orbital period, $P$, is therefore given by:}

\begin{equation}
   P \left(1 - \frac{5^\circ}{360^\circ}\right) \lesssim P_{\rm syn} \lesssim P \left(1 + \frac{5^\circ}{360^\circ}\right)
\end{equation}
\begin{equation}
2.17 {\rm ~days}    \lesssim P_{\rm syn} \lesssim 2.23 {\rm ~days}
\end{equation}

\begin{figure*}[ht!]
    \centering
    	\includegraphics[width=\textwidth]{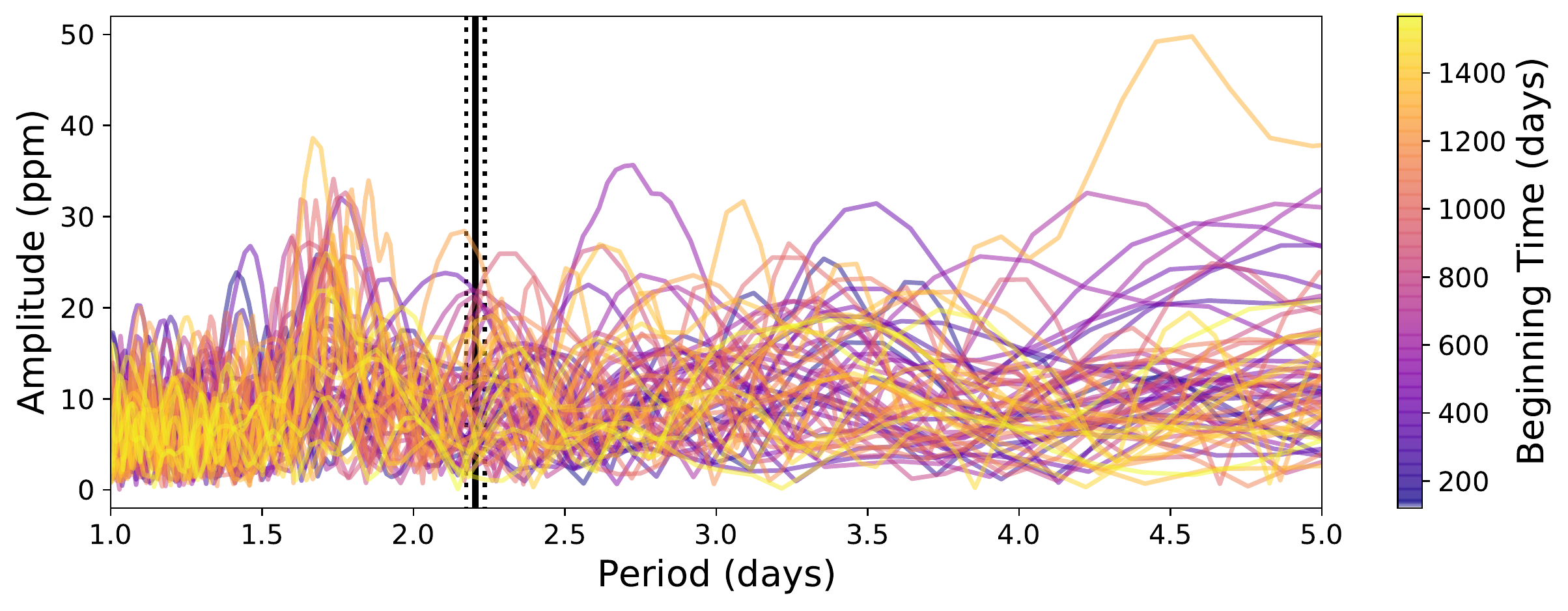}
        \caption{\bmaroon{Lomb-Scargle periodogram of the Kepler short-cadence light curve of \thisstar, with the planetary signal removed. Each line represents one of 60 segments of the light curve. The period of \thisplanet\ is denoted by the vertical black line. The vertical black dotted lines represent the largest range in which we could reasonably expect periodogram features to correspond to variability from the planet itself, based on the fastest phase offset variability we identified in our \thisplanet\ results. There is fairly consistent noise over all periods, but there is not a significant spike near the period of the planet. The implications of this are further discussed in Section \ref{sec:source of variation}.}}
        \label{fig:periodogram}
\end{figure*}

\bmaroon{We plot the Lomb-Scargle periodograms from the individual light curve segments in Figure \ref{fig:periodogram}, and highlight the range of plausible periods at which residual atmospheric variability should leave spikes. We see no clear evidence for any excess noise or peaks in the power spectra near the planet's orbital period, which would have suggested that the phase offset variations was likely due to changes in the planet's atmospheres. Instead, the only features in the power spectra we can see are peaks near 1.7 days that we identify as possible rotation signals, and the broad supergranulation signal that slowly rises towards longer periods. This suggests that the apparent variations we see in \thisplanet's phase curve are likely due to these other astrophysical noise sources. }

        \label{fig:periodogram_and_random}

\subsection{\bmaroon{Possible contributions to the phase offset variability from the periodogram}}
\label{sec:linear_fitting_test}

\bmaroon{We investigate the extent to which the excess noise in \thisstar's light curve can affect our measurement of amplitudes and phase offsets in \thisplanet's phase curve. In Section  \ref{sec:astrophysical noise sources in hatp7b's light curve}, we identified the possible rotation signal at 1.7 days and the broad supergranulation signal in \thisstar's light curve as signals which could plausibly affect our measurement of \thisplanet's phase curve, so we test each of these signals to determine whether they could cause the apparent phase curve variability we see. }

\bmaroon{We devised a test to determine if 
the excess background noise has an influence on our measurement of any given phase offset. We created a fake data set, $f$, to mimic the \thisstar\ light curve, based on a model of a combination of two sinusoids:} 
\begin{equation}\label{fake_dataset_1}
        f = a \cos{\left(\frac{2 \pi t}{P}\right)} + a_n \sin{\left(\frac{2 \pi t}{P_{n}} + \phi_n\right)} + \mathcal{N}(\mu_w, \sigma_w)
\end{equation}

\bmaroon{where $a$ and $P$ are the amplitude and period of a simulated planetary phase curve signal, $a_n$, $P_n$, and $\phi_n$ are the amplitude, period, and phase of a simulated astrophysical noise signal, $t$ is the observation time, and $\mathcal{N}(\mu_w, \sigma_w)$ is a white noise term drawn from a normal distribution with mean $\mu_w$ and standard deviation $\sigma_w$. For our tests, we simulated a planetary phase curve with $a = 35$ ppm (the mean amplitude measured across \thisplanet's phase curves) and $P = 2.204735417$ days, and evaluated the model over a timespan of 22 days of observations spaced like \Kepler\ long cadence data, mimicking the 10-orbit light curve segments we used when measuring the phase curves of \thisplanet. For the white noise term, we set $\mu_w=0$ and $\sigma_w = 25.5714$ ppm, which is equal to the median flux uncertainty in the Kepler light curve.} 

\bmaroon{We tested the impact of two different stellar variability signals: the possible 1.7 day rotation signal, and the supergranulation signal. We used the following parameters to generate simulated light curves for each test:} 
\begin{enumerate}
\item \bmaroon{The possible rotation signal at 1.7 days. For this test, we set the noise period $P_n = 1.7$ days, drew random phases $\phi_n$ from a uniform distribution between 0 and $2\pi$, and drew random amplitudes $a_n$ from a log-normal distribution matching the distribution of amplitudes we measured from the periodograms for of each light curve segment in Section \ref{sec:source of variation} between periods of $1.6$ and $1.9$ days. In detail, the amplitudes (in ppm) were drawn from a distribution described by $10^{\mathcal{N}(1.12, 0.26)}$, and truncated to avoid including amplitudes greater than $40$ ppm. }
\item \bmaroon{The supergranulation signal overlapping \thisplanet's orbital period. For this test, we set the noise period $P_n = 2.204735417$ days (matching the planet's orbital period), drew phases $\phi_n$ from a uniform random distribution between 0 and $2\pi$, and drew random amplitudes $a_n$ from a log-normal distribution matching the distribution of amplitudes we measured from the periodograms for of each light curve segment in Section \ref{sec:source of variation} with periods in the intervals [$2.1,2.15]$ and $[2.25,2.3]$. These period ranges bracket \thisplanet's orbital period, but avoid the synodic period at which we would observe any real variability in \thisplanet's atmosphere (see Section \ref{sec:source of variation}).  The amplitudes (in ppm) were drawn from a distribution described by $10^{\mathcal{N}(0.93, 0.26)}$, and truncated to avoid including amplitudes greater than 25 ppm. Using a single sine component to represent supergranulation, a signal with power over a broad range of frequencies, is a simplification but captures most of the behavior of the supergranulation signal because the other frequency components are orthogonal and do not affect the planet's phase curve as strongly.  }
 \end{enumerate}
 
\bmaroon{For each test, we generated $10,000$ of these simulated light curves and tested how unmodeled stellar variability affects the measured phases and amplitudes by fitting each light curve with an imperfect model, $m$, defined as: }
\begin{equation}\label{imperfect_model}
        m = d\cos{\left(\frac{2 \pi t}{P}\right)} + e\sin{\left(\frac{2 \pi t}{P}\right)}
\end{equation}
\bmaroon{\noindent where $d$ and $e$ are free parameters, and again, $t$ is the observation time and $P$ is \thisplanet's orbital period ($2.204735417$ days). This model ignores the contribution to the simulated light curve from stellar variability (matching the behavior of our MCMC fits in Section \ref{sec:mcmcfitting}). After fitting the simulated light curves with this imperfect model using a linear least squared fitting algorithm \citep{Heiles}, we recovered the best-fit phase, $\phi$ and amplitude, $a$ of the planetary phase curve using:}
\begin{equation}\label{trig2}
        a = \sqrt{d^2 + e^2}
\end{equation}
\begin{equation}\label{trig3}
        \phi = \arctan{\left(-\frac{e}{d}\right)}.
\end{equation}

\bmaroon{We found that the presence of unmodeled variability at the 1.7 day possible rotation period has a relatively small impact on the recovered phase offset and amplitude of \thisplanet's phase curve. The distribution of recovered amplitudes was well described by a normal distribution centered at the input value of 35 ppm, with a standard deviation of only 1.1
ppm. Likewise, the standard deviation of recovered phase offsets was only about $1.9 ~deg$, much smaller than the apparent variations we observe in \thisplanet's phase curve. In hindsight, it is not surprising that the 1.7 day signal does not strongly affect \thisplanet's phase curve parameters, since sinusoids at different frequencies are orthogonal in long time series.}

\begin{figure*}[ht!]
    \centering
    	\includegraphics[width=5.5in]{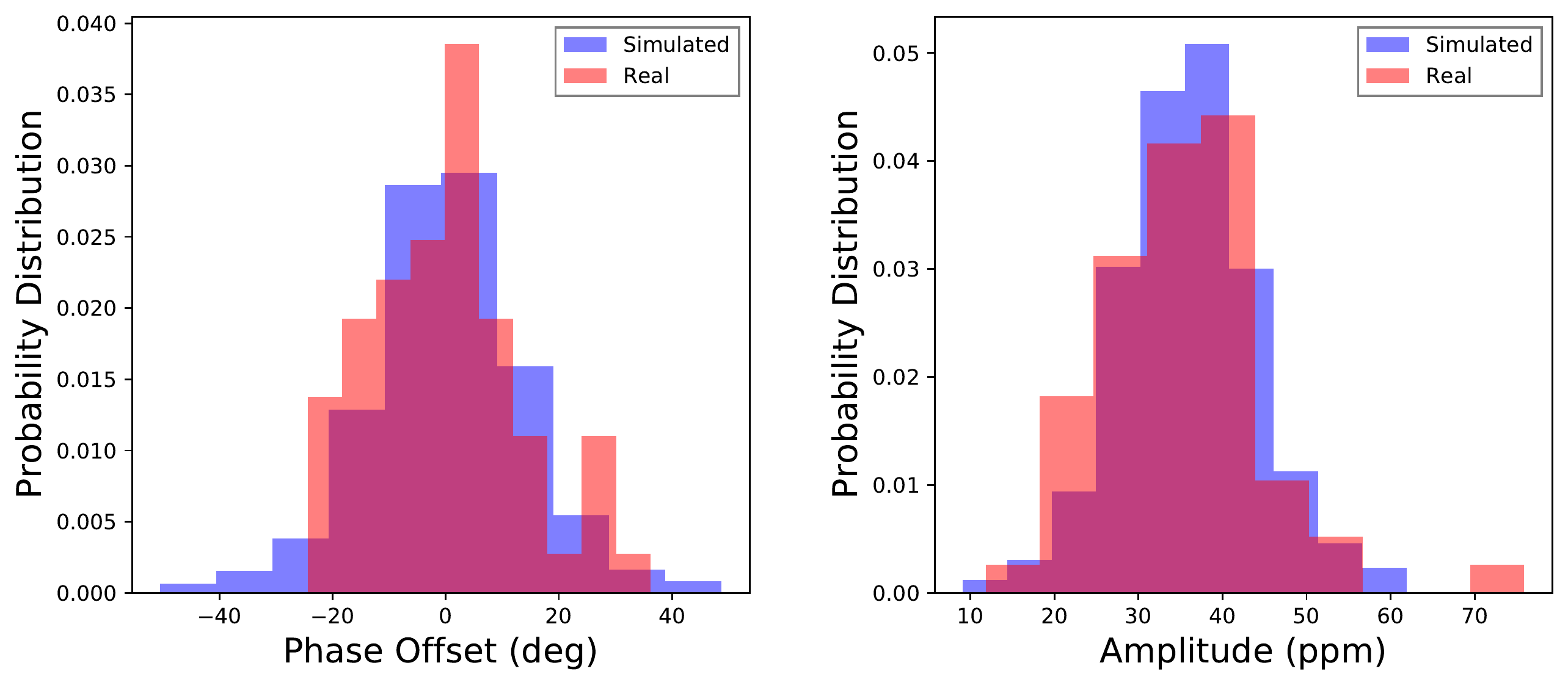}
        \caption{\bmaroon{Comparison of measured distributions and simulated distributions. The red histograms in each panel show the distribution of our phase offset and amplitude variability detected in \thisplanet\ resulting from our original analysis (shown as a time series in Figure \ref{fig:ourresults}). The blue histograms show the results of the test described in section \ref{sec:linear_fitting_test}, wherein a simulated light curve imitating the background noise from the Kepler light curve of \thisplanet\ near the period of the planet was created, and phase offset and amplitude variations were extracted using a linear fitting scheme imitating our initial analysis. The histogram on the left compares our real results to our simulated results for phase offset variations, and the histogram on the right does the same for amplitude variations. These histograms show that the variability we detected from our original analysis can be explained by non-planetary sources of background noise present in the Kepler light curve of \thisplanet, and thus cannot be definitively said to indicate atmospheric variability of the atmosphere of \thisplanet.}}
        \label{fig:sim_histograms}
\end{figure*}

\bmaroon{However, we found that unmodeled variability at the planet's orbital period from the stellar supergranulation signal had a major impact on the measured parameters from \thisplanet's phase curve. The standard deviation of the recovered amplitudes and phase offsets from our simulation was 8.1 ppm and 14.1 deg, respectively -- both significantly larger than the scatter we found in our test of the 1.7 day signal, and much closer to the apparent variations we see in \thisplanet's phase curve (which were 9.9 ppm and 13.3 deg). Figure \ref{fig:sim_histograms} shows a comparison between the distribution of recovered phase offsets and amplitudes from our supergranulation simulations and from our actual measurements of \thisplanet's phase curve. The simulated and actual distributions are close matches, indicating that the unmodeled supergranulation signal in \thisstar's light curve could quite plausibly cause the apparent variations we found in \thisplanet's phase curve. Supergranulation therefore could provide a unified explanation for the phase offset variations and apparently spurious amplitude variations in \thisplanet's phase curve seen by \citet{armstrong2016}. }

\section{Discussion}
\label{sec:discussion}

In this paper, we have shown that the previously claimed variability in the atmosphere of \thisplanet\ may be spurious. We do confirm the presence of apparent variations in the phase and amplitude of \thisplanet's phase curve that at first appear statistically significant. But on further investigation, we find that other processes can also cause apparent variations in the phase and amplitude of non-varying phase curves. 
\bmaroon{We summarize the evidence that the apparent variations in \thisplanet's phase curve are spurious and caused by processes other than atmospheric variability here:
\begin{enumerate}
    \item Injection/recovery tests (Section \ref{sec:injection/recovery}) showed that unchanging phase curve signals injected into similar stars observed by \Kepler\ often show phase curve variations like those we see in \thisplanet. Evidently, it is possible to recover phase curve variations like we see in \thisstar\ even when the underlying signal is perfectly stationary. 
    \item When we remove the average planetary phase curve signal from \thisstar's light curve, we do not see residual peaks near \thisplanet's orbital period in the power spectrum (Section \ref{sec:source of variation}). Any real variability in \thisplanet's atmosphere would leave residual signals when the average phase curve is removed, and these residual signals should be found in the power spectrum very close to the planet's orbital period. We see no evidence for such residual signals in \thisstar's light curve power spectrum. 
    \item Excess noise in \thisstar's light curve from supergranulation can explain the observed phase curve variations. In Section \ref{sec:astrophysical noise sources in hatp7b's light curve}, we showed that \thisstar's photometric variability introduces excess noise into its light curve, and in Section \ref{sec:linear_fitting_test}, we showed that the excess noise from \thisstar's supergranulation naturally explains both the phase offset and amplitude variations in \thisstar's phase curve. 
\end{enumerate}}

\bmaroon{We therefore suspect that the previously reported changes in \thisplanet's phase curve are not due to atmospheric variability, and instead are the result of excess correlated photometric noise from \thisstar's supergranulation. }



\bmaroon{The phenomenon of photometric variability impacting measurements of planetary phase curves} has been investigated previously by \citet{hidalgo}. \bmaroon{These authors find empirically} that stars with effective temperatures between 5500 and 6000 K provide the best phase curve recovery to avoid interference from stellar variability, while hotter stars have less stable variability patterns.  \thisplanet's host star has an effective temperature of 6440 K, making it a more challenging target to reliably measure phase curve variability. 

\bmaroon{While \citet{hidalgo} studied the recovery of phase curve parameters as a function of host star temperature, our results suggest that surface gravity may also be an important parameter for predicting how amenable a given target will be for characterizing atmospheric variability from phase curves. The variability that we measure in \thisplanet's phase curves is likely in large part due to host star supergranulation. The amplitude of photometric variations due to granulation (and likely supergranulation as well) is related to the stellar surface gravity \citep{bastien2013, bugnet}; the lower the surface gravity is in a host star, the higher amplitude granulation and supergranulation signals the star will exhibit. \thisstar\ is a slightly evolved subgiant with lower surface gravity, and therefore higher amplitude photometric variability, than most transiting planet hosts. Future work looking for atmospheric variability from photometric phase curves would target hot Jupiters around stars with higher surface gravity, such as main-sequence Sun-like stars or lower mass hosts like K and M dwarfs.}


Perhaps it is not surprising that we could not confirm the presence of physical variability in the atmosphere of \thisplanet. The result from \citet{armstrong2016} appeared to show very large variations in the phase curve on relatively short timescales, while it is challenging for theoretical models to reproduce such large variability. \citet{komacek} presented hot Jupiter simulations showing that although phase offset in the hottest part of the planet's atmosphere may be observable, those variations are only expected to be as large as $6$ degrees. \citet{rogers} showed that strong magnetic fields can drive winds capable of producing the claimed variability in \thisplanet's atmosphere, but required a magnetic field strength of 6 Gauss, which may be difficult to achieve in a slowly rotating hot Jupiter (although see \citealt{yadav2017} and \citealt{cauley}).   

A detection of atmospheric variability has also been claimed for the planet \Kepler-76 b \citep{jackson}, who performed an analysis similar to that of \citet{armstrong2016}. \citet{jackson} identified variations which were qualitatively similar to those identified in \thisplanet. We have not performed our own analysis of this planet's light curve, but given the similarities between their analysis and results and those on \thisplanet, it is possible that the claimed variability in \Kepler-76 b's atmosphere may also be due to factors other than physical atmospheric variations. Similar to \thisplanet, the host star of \Kepler-76 b has an effective temperature of 6309 K, which is above the ideal range suggested by \citet{hidalgo} to avoid interference from stellar variability. \bmaroon{\Kepler-76 also appears to be a slightly evolved subgiant with lower surface gravity than dwarf stars of the same mass \citep{berger2018, stassuntic}, and therefore may have a large-enough supergranulation signal to cause spurious phase curve variations. Similar analyses to those we performed on \thisstar\ could shed light on the significance of this result. }

Going forward, we recommend using \bmaroon{a similar suite of analyses as performed here, including injection/recovery tests, Fourier analysis, and simulated datasets to test} the significance of any future detection of variability in the atmosphere of an exoplanet. These tests are fairly straightforward to do for wide-field optical surveys like \Kepler\ or \TESS, where many other stars are simultaneously observed and could be used for injection/recovery tests\bmaroon{, and where observations span many orbits and allow for comprehensive characterization of the stellar variability.} In other light curves from \Kepler\ or \TESS, it should be possible to identify similar stars, inject stationary phase curves, and ensure that the statistical significance of the claimed variability is stronger than the strongest spurious signal detected in the injection/recovery tests. \bmaroon{ Likewise, with observations over many orbital periods, it should be possible to separate and characterize the host star's variability in Fourier space. The frequency resolution of a light curve power spectrum increases with the total length of observations; long observational windows like those in \Kepler, and to a lesser extent in K2 and \TESS, make it possible to identify and separate other photometric signals from real changes to the planet's atmosphere. Even though \TESS\ observes for shorter time periods than \Kepler, it also observes in a redder bandpass, where exoplanet phase curves have higher amplitude and stellar variability signals are weaker \citep[e.g.][]{Albregtsen, reiners}.  }

It may be more difficult to perform similar tests to assess the significance of variability detected by targeted telescopes like the \textit{Hubble Space Telescope}, \textit{Spitzer Space Telescope}, or the \textit{James Webb Space Telescope}. In these cases, there will be many fewer datasets suitable for injection/recovery tests, \bmaroon{and these telescopes generally do not perform long-term continuous host star monitoring, so Fourier analysis will be limited by the short time baselines of observations. On the other hand, these observatories will primarily be studying phase curves at redder wavelengths than \Kepler\ and \TESS, and therefore enjoy even lower-amplitude stellar variability. Also, even though \textit{Spitzer} and \textit{JWST} may only observe for a short period of time, by now, \TESS\ has observed most of the sky. It may be beneficial to use \TESS\ to investigate the stellar variability properties of \textit{Spitzer} and \textit{JWST} single-visit targets, to avoid under-reporting uncertainties of phase curve parameters.}


\bmaroon{Finally, it may be possible to circumvent the problem of spurious phase curve variations due to unmodeled stellar variability signals with more sophisticated data analysis methods. In particular, we recommend investigations into modeling correlated variability from supergranulation using Gaussian Process regression. Gaussian Process regression has been successfully applied to a number of astrophysical problems, including modeling stellar variability in \Kepler\ light curves \citep{grunblatt} and radial velocity observations \citep{haywood2014}. Properly accounting for correlated noise in the light curve using a Gaussian Process in our MCMC likelihood function, as opposed to the simple white noise model we used in our analysis, would likely improve our results. Because stellar red noise is present in the in the \Kepler\ lightcurve of \thisstar\ at the period of the planet, it will be difficult for a Gaussian Process to disentangle this correlated noise from the phase curve signal and increase the accuracy of measurements, but the Gaussian Process would likely marginalize over different possible stellar variability signals and avoid underestimating the uncertainty on each data point.} 

\section{Conclusions}
\label{sec:conclusions}
Previously, \citet{armstrong2016} reported the detection of variations in the phase curve of the exoplanet \thisplanet. We attempted to identify and verify atmospheric variability for \thisplanet\ by analyzing the \Kepler\ light curve of the host star. We removed low-frequency variability from the light curve using a basis spline, and used Markov Chain Monte Carlo to fit a phase curve to consecutive time segments of the light curve. We generally confirmed the variability detected by \citet{armstrong2016}, indicating the possible presence of atmospheric variations and cloud movements. We find that under assumptions of pure Gaussian noise, the significance of the detection is extremely high ($p\approx0$), with reduced $\chi^2$ values of 5.4 and \bmaroon{4.8} for phase \bmaroon{offset} and amplitude time series respectively. 

We then tested the robustness of this result. First, we tested whether we measured similar parameters for \thisplanet's phase curve when we changed parts of our methodology. We compared the results using different detrending/low-frequency-removal methods, different \Kepler\ data processing methods, and different models for our MCMC fits. Our results remained generally the same across these different analysis methods, although we did find that our results were fairly sensitive to the particular detrending/low-frequency removal method we used.

We also tested whether other factors besides true atmospheric variability could produce the variability we detected in \thisplanet's phase curve using signal injection/recovery tests. These tests consisted of repeating our phase curve analysis on other stars observed by \Kepler\ into which a non-varying signal similar to \thisplanet's orbital phase curve had been injected. We found that we recovered similarly significant levels of phase offset and amplitude variation from these non-varying sources. This indicated that the variations in the parameters of \thisplanet's phase curve could be caused by processes other than real atmospheric variations, \bmaroon{and the result required closer analysis.}

\bmaroon{We then investigated what other factors besides true atmospheric variations could cause the variations we measure. We removed the planetary signal from the \Kepler\ light curve and calculated periodograms of the full light curve and shorter light curve segments. We found high levels photometric variability (in excess of random noise) due to supergranulation on timescales close to the orbital period of \thisplanet, as well as a possible stellar rotation period at $1.7$ days. We used simulated 10-orbit light curves with the same level of noise as the \Kepler\ data, including proxies for these variability signals, and tried to recover phase curve parameters using a model ignoring the variability. These tests showed that the possible rotation signal at $1.7$ days did not have a significant effect on our phase curve measurements, but the supergranulation noise near the period of \thisplanet\ did introduce apparent phase offset and amplitude variability on the scale of what we measured for \thisplanet. This indicates that unmodeled stellar noise from supergranulation could be entirely responsible for the variability we measure in the phase offset and amplitude of \thisplanet.}


This result underscores the extreme difficulty of robustly measuring variability in the atmospheres of exoplanets. With four years of extremely precise photometry (60 ppm per 30 minute exposure), the \Kepler\ dataset on \thisstar\ is one of the highest-quality photometric time series in existence. This work suggests that future observers will need to consider the host star's variability before characterizing the variability of the planet's atmosphere.   

\acknowledgments

We acknowledge support from the UT Austin REU grant AST-1757983 (PI: Jogee) funded by the NSF REU and DOD ASSURE programs. We thank Shardha Jogee, Avi Shporer, and Thaddeus Komacek for comments on the manuscript and Jeff Smith, Dan Huber, and George Zhou for helpful conversations. We especially thank David Armstrong for useful comments and conversations on the manuscript that strengthened and sharpened our analysis, and for suggesting the layout of Figure \ref{fig:excess_noise}. We thank the anonymous referee for a constructive report that significantly improved the paper. AV's work was partially performed under contract with the California Institute of Technology (Caltech)/Jet Propulsion Laboratory (JPL) funded by NASA through the Sagan Fellowship Program executed by the NASA Exoplanet Science Institute. This research has made use of NASA's Astrophysics Data System and the NASA Exoplanet Archive, which is operated by the California Institute of Technology, under contract with the National Aeronautics and Space Administration under the Exoplanet Exploration Program. Some of the data presented in this paper were obtained from the Mikulski Archive for Space Telescopes (MAST). STScI is operated by the Association of Universities for Research in Astronomy, Inc., under NASA contract NAS5--26555. Support for MAST for non--HST data is provided by the NASA Office of Space Science via grant NNX13AC07G and by other grants and contracts. This paper includes data collected by the \Kepler\ mission. Funding for the \Kepler\ mission is provided by the NASA Science Mission directorate.

%

\vspace{5mm}
\facilities{\textit{Kepler}}


\software{numpy \citep{np}, matplotlib \citep{plt}, Astropy \citep{astropy:2018}, scipy \citep{2020SciPy-NMeth}, lightkurve \citep{lightkurve}}



\clearpage



\end{document}